\documentclass[english,10pt,aps,pra,twocolumn,superscriptaddress,floatfix,fleqn]{revtex4-2}
\pdfoutput=1
\renewcommand{\bibliography}[1]{}

\usepackage[utf8]{inputenc}
\usepackage[T1]{fontenc}
\usepackage{amsmath}
\usepackage{amssymb}
\usepackage{amsfonts}
\usepackage{bm}
\usepackage{bbm}
\usepackage{epsfig}
\usepackage{grffile}
\usepackage{times}

\usepackage[usenames,dvipsnames]{color}
\definecolor{dblue}{rgb}{0,0.1,.6}
\definecolor{dred}{rgb}{.6,0.1,0}

\usepackage[colorlinks=true,citecolor=dblue,linkcolor=dblue,urlcolor=dblue]{hyperref}
\usepackage[all]{hypcap}

\usepackage{enumitem}
\setlist[enumerate]{label=(\alph*), itemsep=0pt, leftmargin=*}
\newcommand{\ud}{\mathrm{d}}
\newcommand{\mc}[1]{\mathcal{#1}}

\renewcommand{\vec}[1]{{\boldsymbol{#1}}}

\newcommand{\A} {\mc{A}}
\newcommand{\C} {\mc{C}}
\newcommand{\E} {\mc{E}}
\newcommand{\G} {\mc{G}}
\newcommand{\K} {\mc{K}}
\renewcommand{\O} {\mc{O}}

\newcommand{\W} {\mc{W}}
\newcommand{\vg}{\vec{g}}
\newcommand{\vs}{\vec{\sigma}}
\newcommand{\vx}{\vec{x}}
\newcommand{\vy}{\vec{y}}
\newcommand{\vz}{\vec{z}}

\newcommand{\Np}{\mc{N}}
\newcommand{\Nb}{\mc{M}}
\newcommand{\sump}{\sideset{}{'}\sum}
\newcommand{\const}{\textnormal{const}}
\newcommand{\bg}{\textnormal{bg}}
\newcommand{\tR}{\textnormal{R}}
\newcommand{\tI}{\textnormal{I}}
\newcommand{\tS}{\textnormal{S}}
\newcommand{\ta}{\textnormal{a}}
\newcommand{\bta}{\bar{\textnormal{a}}}
\newcommand{\taa}{\textnormal{aa}}
\newcommand{\swap}{\textnormal{swap}}
\newcommand{\lock}{\textnormal{locked}}

\newcommand{\Emph}[1]{\emph{\textbf{#1}}}
\let\up\uparrow
\let\dw\downarrow

\newcommand  {\Pmatrix}[1]{\begin{pmatrix}#1\end{pmatrix}}

\usepackage{amsthm}
\newtheorem{example}{Example}

\newcommand{\qlab}  {National Quantum Laboratory, University of Maryland, College Park, MD 20742, USA}
\newcommand{\umd}   {Department of Physics, University of Maryland, College Park, MD 20742, USA}
\newcommand{\duke}  {Department of Physics, Duke University, Durham, North Carolina 27708, USA}

\begin{document}

\title{Conditional-path Monte Carlo for rare stochastic dynamics on networks: Details and derivations}
\author{Thomas Barthel}
\affiliation{\umd}
\affiliation{\qlab}
\affiliation{\duke}
\author{Jiazheng Sun}
\affiliation{\duke}
\author{Jhao-Hong Peng}
\affiliation{\duke}
\date{August 8, 2026}

\begin{abstract}
The simulation of rare macroscopic events in stochastic network dynamics, such as widespread epidemic outbreaks, cascading failures in communication networks, or the escape from metastable states in many-body systems, is severely hindered by methodological challenges like catastrophic rejection rates, weight degeneracy, genealogical correlations, and critical slowing down inherent to standard forward-time algorithms, splitting methods, and transition-path sampling. Conditional-path Monte Carlo (CPMC) overcomes these limitations by employing non-local Swendsen-Wang-like cluster updates that operate directly on full-system trajectories. Serving as the technical companion to [Sun, Moody, and Barthel, arXiv:2608.16171], this paper provides the rigorous mathematical foundations and algorithmic details underlying the CPMC framework. We formally define the joint path-graph probability weights and derive the transition and uniformization sum rules that guarantee detailed balance. Applying the framework to susceptible-infectious-susceptible (SIS) models, we systematically construct and optimize single-node and edge graph vertex sets specifically designed to prevent lock avalanches and maintain the structural mobility of the epidemic trunk. Furthermore, we detail a dynamic programming scheme to exactly implement complex boundary conditions -- including patient-zero and macroscopic outbreak-size constraints -- enabling the rejection-free generation of valid trajectories. Finally, we assess the computational complexity of the algorithm, describe parallelization strategies, and validate CPMC against exact solutions for dynamics on small networks.
\end{abstract}

\maketitle

\section{Introduction}\label{sec:intro}
Understanding the stochastic evolution of complex networks and many-body systems is a central challenge across physics, biology, engineering, social science, and finance. In many real-world systems, the most consequential macroscopic events emerge from a rare confluence of localized stochastic processes. In statistical mechanics, for instance, rare extreme fluctuations drive the escape from metastable states \cite{Haenggi1990-62}, such as nucleation in kinetic spin models \cite{Binder1987-50,Rikvold1994-49}, and dictate the evolution of glassy systems \cite{Garrahan2007-98,Chandler2010-61}. For infectious diseases, only a small percentage of local surges expand to trigger widespread epidemics because they must navigate specific bottlenecks in the contact patterns of heterogeneous populations \cite{PastorSatorras2015-87,LloydSmith2005-438,Salathe2010-6}. Similarly, cascading failures in communication networks and financial systems often originate from highly improbable combinations of local faults \cite{Albert2000-406,Acemoglu2015-105}, and massive shifts in societal opinion can be driven by a rare alignment of local influences \cite{Castellano2009-81,Watts2002-99}. Because these pivotal events are inherently rare, analyzing them with direct forward-time simulations is generally inefficient.

Traditional individual-based simulation techniques of stochastic network dynamics like the susceptible-infectious-susceptible (SIS) model \cite{Hethcote1989,Hethcote2000-42,Brauer2017-2,Rock2014-77,Brauer2019} are forward-time algorithms -- most notably the \emph{Gillespie method} \cite{Gillespie1977-81} and the intimately related \emph{kinetic Monte Carlo} \cite{Bortz1975-17}. These standard \emph{stochastic simulation algorithms} (SSA) propagate an initial state forward in time by randomly sampling the next elementary state transition according to the physical rates of the system. While SSA is exact and highly efficient for exploring typical, unconstrained dynamics, it suffers from \emph{catastrophic rejection rates} when simulating rare events. Attempting to simulate rare events that are formulated in terms of macroscopic constraints (like a specific outbreak-size threshold) requires discarding the vast majority of generated paths that fail to satisfy the constraint.

To circumvent the massive rejection rates of forward-time SSA, various valuable rare-event sampling techniques have been developed:
\begin{itemize}
 \item
 The \emph{weighted stochastic simulation algorithm} (wSSA) \cite{Kuwahara2008-129,Gillespie2009-130}, developed to study rare biochemical events, relies on importance sampling. It modifies the standard Gillespie algorithm by artificially biasing local reaction propensities to actively drive trajectories toward a rare target state, then unbiasing the final probability estimate using the likelihood ratio with respect to the true model. For large stochastic networks, wSSA faces practical challenges: In complex topologies, an optimal selection of heuristic biasing parameters is difficult, and suboptimal choices can drastically inflate the estimator's variance. Furthermore, as network size or simulation time increases, wSSA suffers from \emph{weight degeneracy}, where a tiny fraction of trajectories accumulates massive statistical weight and dominates the ensemble, effectively destroying the sample size. Finally, wSSA is still a local, forward-time integrator. When it must navigate rigid topological bottlenecks step by step, it is challenged by \emph{kinetic trapping} \footnote{Imagine dense communities connected by only a few bridging edges and nodes. For an epidemic to grow from a localized outbreak in a small community into a massive, system-wide epidemic, the infection must cross the network bridges. If the transmission rate across the bridges are low, they become rigid topological bottlenecks. Forward-time algorithms like wSSA, FFS, and WE are blind to the future. Even if wSSA artificially biases the rates to encourage more infections, the algorithm still has to generate precarious sequences of events, where infections reach bridges and cross them.}.
 \item
 \emph{Splitting methods}, such as \emph{forward flux sampling} (FFS) \cite{Allen2006-124a,Allen2006-124b,Allen2009-21} or \emph{repetitive simulation trials after reaching thresholds} (RESTART) \cite{VillenAltamirano1991-15,VillenAltamirano1994,VillenAltamirano2002-13}, define an effective order parameter $\lambda$, evolve trial trajectories with SSA, and branch (clone) trajectories that successfully cross intermediate thresholds $\lambda_0<\lambda_1<\lambda_2<\dotsc$ toward the rare event of interest \footnote{FFS and RESTART both rely on defining an order parameter $\lambda$ or importance function and place a series of increasing thresholds or interfaces between the initial state and the rare target state. They were developed in different academic fields -- RESTART in telecommunications/queuing theory and FFS in statistical mechanics -- and use distinct schemes for the branching and termination of trajectories. To proceed from threshold to threshold, FFS uses SSA to evolve numerous stored trial trajectories from $\lambda$ and tracks the fraction of these trajectories that reach the next interface $\lambda_{n+1}$. In this way, one obtains transition probabilities $P(\lambda_{n+1}|\lambda_{n})$ and trial trajectories at $\lambda_{n+1}$. If $\lambda$ is a discrete variable, like the number of infected nodes, some modifications are necessary. In RESTART, trajectories evolve with SSA and branch into $r_n$ clones when crossing $\lambda_n$. If a clone later falls again below $\lambda_n$, it is terminated.}. While powerful, splitting methods can suffer heavily from \emph{path degeneracy} in large, heterogeneous networks. A simple order parameter $\lambda_n$ like the total number of infected nodes generally masks hidden barriers. For example, 10 infected nodes clustered in a dense core have a vastly different probability of driving a massive outbreak compared to 10 infected nodes scattered at the periphery. Using SSA, splitting methods are generally also hampered by \emph{critical slowing down} as trajectories need to ``diffuse'' into rarer states to pass from $\lambda_n$ to $\lambda_{n+1}$. Furthermore, due to the cloning, splitting methods are vulnerable to \emph{genealogical correlation} (the ``founder effect''): When the forward-time dynamics become trapped in topological bottlenecks or slow modes, the algorithm may simply replicate highly correlated offspring \cite{Allen2009-21,Zuckerman2017-46,Aristoff2018-52}.
 \item
 The \emph{weighted ensemble} (WE) algorithm \cite{Huber1996-70,Rojnuckarin1998-95,Zhang2010-132,Donovan2013-139,Zuckerman2017-46}, developed in the biophysics community, is a conceptually somewhat different splitting method. Using SSA, it propagates a population of trajectories (``walkers'') for short time intervals and, similar to the above, maintains order-parameter bins $[\lambda_n,\lambda_{n+1})$. When reaching the end of a time interval, trajectories that have advanced into rarely visited bins are replicated (split) to enhance the exploration of the rare transition, while overpopulated bins are culled to prevent a combinatorial explosion of paths. Applications of WE to large stochastic networks can again be challenged by the path degeneracy problem, critical slowing down, and genealogical correlations.
 \item
 \emph{Transition path sampling} (TPS) \cite{Dellago1998-108,Dellago1998-108b,Bolhuis2002-53,Dellago2002-1,Bolhuis2021-4}, developed for the study of rare transitions between long-lived states in chemical reactions and protein folding, operates directly in the space of trajectories, proposing new valid paths via local ``shooting'' or ``shifting'' moves \footnote{The typical procedure in TPS is to start with an initial (often artificially constructed or high-energy) trajectory connecting a $t=0$ state of class $\mc{A}$ to $t=T$ state of class $\mc{B}$. In the shooting move, one selects a random time slice $t$ along the path, perturbs the microscopic momenta (or variables) slightly, and integrates the equations of motion forward and backward in time to generate a new trial trajectory. If the new trial path successfully connects an $\mc{A}$ state to a $\mc{B}$ state, it is accepted or rejected based on a Metropolis criterion that guarantees preservation of the true path probability distribution.}. However, in constrained network models, such local trajectory updates routinely suffer from critical slowing down, as the algorithm struggles to substantially mutate the topological core of the trajectory (e.g., the trunk of epidemic infection trees) without violating physical rules.
\end{itemize}

To overcome the limitations of forward-time SSA, splitting methods, and local path updates, we introduce the \emph{conditional-path Monte Carlo} (CPMC) \cite{Sun2026_08}. Building on ideas from computational condensed-matter physics, CPMC operates directly on the space of full-system trajectories. Instead of attempting to propose local, step-by-step changes, CPMC maps the current trajectory of the full system to an intermediate graph configuration that decomposes the entire spacetime volume into connected clusters. By employing Swendsen-Wang-like cluster updates, the algorithm simultaneously flips the states of non-local spacetime regions. This allows CPMC to generate a Markov chain of trajectories that all strictly respect the desired macroscopic constraints, entirely bypassing the massive rejection rates and critical slowing down that plague SSA-based methods and TPS in various applications, as well as the path degeneracy problem and genealogical correlations in the splitting methods.

This manuscript serves as the technical companion to Ref.~\cite{Sun2026_08}, which introduced the broader CPMC framework and presented a numerical analysis for large outbreaks in SIS dynamics on kinship networks. Here, we provide the rigorous mathematical foundations and detailed derivations underlying the algorithm. Following a short account of the SIS model in Sec.~\ref{sec:sis_model}, Sec.~\ref{sec:pathProb} introduces important notations and the path probability density. Section~\ref{sec:clusterUpdate} describes the central idea of the Swendsen-Wang-like cluster updates, randomly assigning graph vertices to a trajectory, identifying free spacetime clusters, and randomly selecting the states of these clusters to obtain a new trajectory. We prove detailed balance of the CPMC algorithm. The detailed balance relies on certain sum rules that the insertion rates of the graph vertices need to obey. For SIS dynamics, we discuss different sets of graph vertices and solutions for the associated rate sum rules in Sec.~\ref{sec:SIS-vertexSets}. Again for the example of SIS models, Sec.~\ref{sec:BC} demonstrates how trajectory constraints like patient-zero and outbreak-threshold conditions can be implemented exactly and efficiently during cluster updates through dynamic programming tables. The scaling of computation costs and parallelization strategies are assessed in Secs.~\ref{sec:costs} and \ref{sec:parallel}. We conclude with validation simulations, comparing CPMC to exact solutions on small networks in Sec.~\ref{sec:validate}, and the discussion section~\ref{sec:discuss}.

\section{SIS Model}\label{sec:sis_model}
While the method is applicable for general Markovian stochastic dynamics, for instructive examples, we will exclusively employ the SIS model \cite{Hethcote1989,Hethcote2000-42,Brauer2017-2,Rock2014-77,Brauer2019}, where $N$ individuals form the nodes of an interaction network. A node $i$ can be in two states: infected ($\sigma_i=1$) or susceptible ($\sigma_i=0$). An infected node $i$ spreads the disease to a susceptible node $j$ at \emph{infection rate} $\alpha_{i,j}$, defining the network structure. Additionally, an infected node $i$ returns to the susceptible state at the \emph{recovery rate} $\gamma_i$. For all examples discussed in this work, we consider the continuous-time SIS model where, in general, $\alpha_{i,j}\neq\alpha_{j,i}$ and all rates can depend on the event position (node or edge).

\section{Path probabilities and notations}\label{sec:pathProb}
The \emph{system state} at time $t$ is denoted by
\begin{equation}
	\vs^t:=(\sigma^t_1,\sigma^t_2,\dotsc,\sigma^t_N)\in d^N
\end{equation}
with single-node states $\sigma^t_i\in\{0,\dotsc,d-1\}$. We use $\vs^t_\vx=(\sigma^t_{x_1},\dotsc,\sigma^t_{x_n})$ to denote the state of nodes $\vx=(x_1,\dotsc,x_n)$ at time $t$, and a \emph{trajectory} $(\vs^t\mid 0\leq t\leq T)$ for the evolution from $t=0$ to $T$ is denoted by $\vs\in[0,T]\times d^N$.

Let $\E_\vx^t=\{\vs_x\}$ be the set of all possible elementary \emph{state changes}
\begin{equation}\label{eq:def-stateChange}
	\vs_x:=(\vs_\vx^{t-}\to\vs_\vx^{t+})
\end{equation}
at time $t$ and position $\vx$ in the network as defined by the physical model, where we have introduced the short-hand notation $x=(t,\vx)$ for spacetime \emph{locations}, \emph{positions} $\vx$ are ordered lists of the involved nodes, $\vs_\vx^{t-}$ is the state on these nodes just before the transition, and $\vs_\vx^{t+}$ is the state just after the transition. The associated elementary \emph{event rates} are denoted by $\omega_x(\vs_x)$.

\begin{example}
In the SIS model, node $i$ recovering at time $t$ from an infection corresponds to the state change
\begin{subequations}
\begin{equation}\label{eq:recover}
	\vs^{\tR}_x=(\sigma^{t-}_i\to \sigma^{t+}_i)=(1\to 0)
\end{equation}
at location $x=(t,i)$, which happens at event rate $\omega_x(\vs^{\tR}_x)=\gamma_i$. 
Infections between nodes $i$ and $j$ at time $t$ cause the state changes
\begin{align}\label{eq:infect-up}
	\vs^{\tI\up}_x &=(\sigma^{t-}_i,\sigma^{t-}_j\to \sigma^{t+}_i,\sigma^{t+}_j)=(1,0\to 1,1),\\
	\vs^{\tI\dw}_x &=(\sigma^{t-}_i,\sigma^{t-}_j\to \sigma^{t+}_i,\sigma^{t+}_j)=(0,1\to 1,1)
\end{align}
\end{subequations}
at location $x=(t,i,j)$, where $i<j$ for some arbitrary node ordering. They occur at rates $\omega_x(\vs^{\tI\up}_x)=\alpha_{i,j}$ and $\omega_x(\vs^{\tI\dw}_x)=\alpha_{j,i}$.
Accordingly, the sets of elementary state changes are $\E_i^t=\{\vs^{\tR}_{(t,i)}\}$ and $\E_{i,j}^t=\{\vs^{\tI\up}_{(t,i,j)},\vs^{\tI\dw}_{(t,i,j)}\}$.
\end{example}

Valid trajectories $\vs$ are equivalently specified by an initial state $\vs^0$ and the sequence $\big((x_1,\vs_{x_1}),\dotsc,(x_\Np,\vs_{x_\Np})\big)$ of transition events.

Under trajectory constraints denoted by an indicator function $C(\vs)$, the (unnormalized) \emph{path probability density} is given by
\begin{equation}\label{eq:def-P}
	P(\vs) = C(\vs) \Big(\prod_{x\in\vs}\omega_x(\vs_x)\Big)e^{-\int_0^T\ud t\,\Lambda^t(\vs^t)},
\end{equation}
where the product runs over all locations $x$ of transition events in the trajectory $\vs$.
In a typical scenario, $C$ only depends on the states $\vs^0$ and $\vs^T$ at the start and end times. More generally, $C$ can be a probability distribution for initial and final states, one can include constraints at intermediate times, and work with time-correlated constraints.

\begin{example}
In the SIS model, an infection event \eqref{eq:infect-up} at location $x=(t,i,j)$ contributes the factor $\omega_x(\vs_x)=\alpha_{i,j}$ in Eq.~\eqref{eq:def-P}. A recovery event \eqref{eq:recover} at location $x=(t,i)$ contributes the factor $\omega_x(\vs_x)=\gamma_i$. The indicator function $C$ may encode trajectory constraints such as having a single infected node at time $t=0$ and an outbreak with $\geq N/4$ infected nodes at time $t=T$. In this case, $C(\vs)=C_0(\vs^0)C_T(\vs^T)=\delta_{\sum_i\sigma_i^0,1}\theta(\sum_i\sigma_i^T-N/4)$, where $\theta$ is the Heaviside step function.
\end{example}

In Eq.~\eqref{eq:def-P}, $\Lambda^t(\vs^t)$ is the \emph{total escape rate} of the system from state $\vs^t$ at time $t$. Given the sets $\E_\vx^t$ of possible elementary state changes \eqref{eq:def-stateChange} at time $t$ and position $\vx$,
\begin{subequations}\label{eq:def-Lambda}
\begin{gather}\label{eq:def-Lambda-total}
	\Lambda^t(\vs^t)=\sum_\vx \Lambda_x(\vs_\vx^t)\quad\text{with}\\
	\Lambda_x(\vs_\vx^t):=\sum_{\tilde{\vs}_x\in\E_\vx^t}\delta_{\tilde{\vs}_\vx^{t-},\vs_\vx^t}\,\omega_x(\tilde{\vs}_x).
\end{gather}
\end{subequations}

\begin{figure*}[t]
\label{fig:vertices1}
\centering
\includegraphics[width=0.93\textwidth]{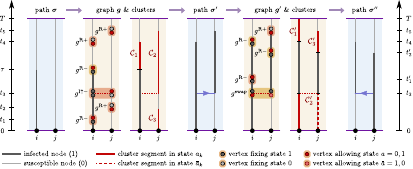}
\caption{\textbf{Two cluster updates for a simple SIS trajectory.} The cluster updates for two nodes of an SIS model described in Example~\ref{ex:vertices1} start from a trajectory $\vs$, where node $i$ is infected until time $\tau$, and node $j$ is susceptible for the entire time $[0,T]$. In the construction of graph $\vg$, the recovery of node $i$ is matched by the vertex $g^{\tR-}_{(\tau,i)}=(1\to a)$, where state $a$ is free to take values 0 (susceptible) and 1 (infected). Additional four single-node background vertices and one edge background vertex are inserted by Poisson point processes; indicated by a redder background color. The edge vertex $g^{\tI\up-}_{(t_3,i,j)}=(1,0\to 1,a)$ fixes node $i$ to be infected before and after time $t_3$, hence fixing the free $t_1^+$ state of the vertex $g^{\tR-}_{(t_1,i)}$. In conjunction with the vertex $g^{\tR+}_{(t_5,j)}=(a\to 0)$, the edge vertex leads to the free cluster $\C_2$ for times $[t_3,t_5]$ on node $j$. Vertices $g^{\tR-}_{(\tau,i)}$ and $g^{\tR+}_{(t_4,i)}$ lead to the free cluster $\C_1$ for times $[\tau,t_4]$ on node $i$, and we have another free cluster $\C_3$ on node $j$. Choosing the infected state for clusters $\C_1$ and $\C_2$ as well as the susceptible state for $\C_3$, we obtain the new trajectory $\vs'$ with node $j$ getting infected by node $i$ at time $t_3$. Another cluster update, where the infection event is matched by the vertex $g^\swap_{t_3,i,j}=(a,\bar{a}\to 1,1)$ and $\bar{a}$ is the negation of $a$, causes a change of the infection direction, moving patient zero from node $i$ to node $j$. It also moves the recovery time of node $j$ from $t_5$ to $t_2'$.}
\end{figure*}

\begin{example}\label{ex:trajectory1}
For the SIS model, we have
\begin{equation}
	\Lambda^t(\vs^t)=\sum_{i\,:\,\sigma^t_i=1} \gamma_i + \sum_{i\,:\,\sigma^t_i=1} \sum_{j\in\partial_i\,:\,\sigma^t_j=0} \alpha_{i,j},
\end{equation}
where $\partial_i$ is the set of node $i$'s nearest neighbors, such that
\begin{equation}
	\int_0^T\ud t\,\Lambda^t(\vs^t)=\sum_i \gamma_i T^\tI_i+\sum_i\sum_{j\in\partial_i} \alpha_{i,j} T^{\tI\tS}_{i,j},
\end{equation}
where $T^\tI_i$ is the total infected time of node $i$, and $T^{\tI\tS}_{i,j}$ is the total time for which node $j$ was susceptible while node $i$ was infected.

An SIS trajectory from time $0$ to time $T$, where node $i$ is infected from time $0$ to time $\tau$ and all other nodes are susceptible as shown in Fig.~\ref{fig:vertices1}, hence has the probability density
$C(\vs) \gamma_i\, e^{-\gamma_i \tau -\sum_{j\in\partial_i}\alpha_{i,j}\tau}$.
\end{example}

\section{Swendsen-Wang-like cluster updates}\label{sec:clusterUpdate}
To efficiently explore the path space under the constraint $C$, CPMC generates a Markov chain of trajectories
\begin{equation}\label{eq:pathChain}
	\vs[1]\xrightarrow{\W} \vs[2]\xrightarrow{\W} \vs[3]\xrightarrow{\W} \dotsc
\end{equation}
all respecting the constraint, where the \emph{path transition probability density} $\W(\vs'|\vs)$ for trajectory updates obeys the detailed balance condition
\begin{equation}\label{eq:detail_balance}
	\W(\vs'|\vs)P(\vs)=\W(\vs|\vs')P(\vs').
\end{equation}

\subsection{Structure of cluster updates and notations}\label{sec:clusterUpdateIntro}
As local trajectory updates are often plagued by critical slowing down, we employ non-local cluster updates similar to the Swendsen-Wang algorithm from computational condensed matter physics and statistical mechanics \cite{Swendsen1987-58,Edwards1988-38}, which has been extensively used for the investigation of classical and quantum many-particle systems in thermal equilibrium \cite{Evertz1993-70,Sandvik1999-59,Evertz2003-52,Krauth2006}.

The cluster update introduces an intermediate \emph{graph} configuration $\vg$, splitting the trajectory update into two steps
\begin{equation}\label{eq:clusterUpdate-sgs}
	\vs \xrightarrow{(a)} \vg \xrightarrow{(b)} \vs'.
\end{equation}
\begin{enumerate}
 \item[(a)]
 In the first step, we translate the given trajectory $\vs$ to a compatible graph $\vg$. Multiple graphs are compatible with a given trajectory $\vs$; the graph choice is random according to a detailed-balance condition in the enlarged state space. A graph $\vg$ provides a decomposition of spacetime $[0,T]\times\{1,\dotsc,N\}$ into clusters $\{\C_k\}$. Each cluster $\C_k$ is characterized by a \emph{base state} $a_k$ which is free to take one of $d_k\leq d$ values, where $d=2$ in the SIS model. A cluster is called \emph{fixed} if $d_k=1$ and \emph{free} if $d_k\geq 2$. Every cluster consists of time intervals on individual nodes $\{i\}$, where the node state $\sigma_i^t$ on each time segment is constant and determined by an injective function $a_k\mapsto \sigma_i^t(a_k)\in\{0,\dotsc,d-1\}$, including the segment's state in the original trajectory $\vs$.
 In this way, graphs $\vg$ define (non-disjoint) finite subsets of trajectories, and every valid trajectory is contained in a whole continuum of graphs as will become clear in the following.
 \item[(b)]
 In the second step, we randomly choose one of the $d_k$ allowed states for each free cluster $\C_k$ with uniform probability $1/d_k$ such that we obtain a new trajectory $\vs'$.
\end{enumerate}

The graph $\vg$ consists of so-called \emph{graph vertices} $g_z$ at locations $z$. Each vertex imposes a certain local constraint on the trajectories that are deemed compatible with the graph, restricting allowed states for the $t^-$ and $t^+$ states of the involved nodes or imposing bijective relations between these states.
Specifically, the graph $\vg$, constructed in step (a) of the update, comprises one graph vertex $g_x$ for every event location $x\in\vs$, corresponding to the $\Np_\vs$ \emph{physical} state changes $\vs_x$, and further $\Nb_{\vg,\vs}$ \emph{background} vertices $g_y$ inserted via a Poisson point process at locations $y$ on time intervals where no physical change occurs in $\vs$.

For brevity of notation we write 
\begin{itemize}
 \item ``$\vs\in\vg$'' for all trajectories $\vs$ that are compatible with graph $\vg$,
 \item ``$\vg\ni\vs$'' for all graphs $\vg$ that are compatible with trajectory $\vs$, i.e., the graphs that contain $\vs$,
 \item ``$x\in\vs$'' for the locations of all $\Np_\vs$ state-changing events in trajectory $\vs$, and
 \item ``$z\in\vg$'' for the locations of all $\Np_\vs+\Nb_{\vg,\vs}$ graph vertices in $\vg$.
\end{itemize}
As described above, for a given trajectory $\vs\in\vg$, we can arrange the vertices of graph $\vg$ into a group of $\Np_\vs$ physical vertices $\{(x,g_x)\}$ and a group of $\Nb_{\vg,\vs}$ background vertices $\{(y,g_y)\}$.
Note that for different trajectories $\vs,\vs'\in\vg$, a vertex at $z\in\vg$ can be physical with respect to $\vs$ but a background vertex with respect to $\vs'$. In fact, the nature of graph vertices can change in step (b) of the cluster update from physical to background or vice versa.

Analogous to the sets $\E_\vx^t$ of elementary state changes \eqref{eq:def-stateChange}, we introduce the sets of possible graph vertices at time $t$ and position $\vz$, denoted by
\begin{equation}\label{eq:def-graphVertices}
	\G_z\equiv\G_\vz^t=\{g_z\}\quad\text{with}\quad z=(t,\vz).
\end{equation}

\begin{example}\label{ex:vertices1}
Consider the trajectory $\vs$ from Example~\ref{ex:trajectory1}, where $x=(\tau,i)$ is the location of a recovery event in the trajectory $\vs$ for an SIS model. Two physical graph vertices for location $x$ that are compatible with $\vs_x$ are (a) a vertex $g^{\tR-}_x=(1\to a)$ that enforces the infected state $\sigma^{\tau-}_i=1$ just before time $t=\tau$ but allows any state $\sigma^{\tau+}_i=a\in\{0,1\}$ after time $\tau$ and (b) a vertex $g^{\tR+}_x=(a\to 0)$ that allows any state $\sigma^{\tau-}_i=a\in\{0,1\}$ just before time $\tau$ but enforces the susceptible state $\sigma^{\tau+}_i=0$ after time $\tau$. A vertex of type (a) could also be inserted as a background vertex $g^{\tR-}_y$ at any location $y=(t,i)$ with $0<t<\tau$. Similarly, a vertex of type (b) could also be inserted as a background vertex $g^{\tR+}_y$ at any location $y=(t,i)$ with $\tau<t<T$. See Fig.~\ref{fig:vertices1}.

In addition to these single-node vertices for recovery events in the SIS model, we employ further single-edge vertices for infection events as detailed in Sec.~\ref{sec:SIS-vertexSets}. One example is the vertex $g^{\tI\up-}_y=(1,0\to 1,a)$, which we can insert as a background vertex at any location $y=(\tau',i,j)$ with $\tau'<\tau$. The current trajectory $\vs$ corresponds to $a=0$. Changing $a$ to 1 in a cluster update introduces an infection event from node $i$ to node $j$ at location $y$ of the new trajectory $\vs'$. Now, in the next cluster update, we can for example assign the vertex $g^\swap_x=(a,\bar{a}\to 1,1)$ to the infection event, where $a=1$ for trajectory $\vs'$ and $\bar{a}$ denotes the negation of $a$. If no other vertices fix the states of nodes $i$ and $j$ for time $\tau^{\prime-}$, the cluster update can change the variable $a$ to 0, effectively switching the infection source such that, in the new trajectory $\vs''$, node $j$ infects node $i$ at time $\tau'$, and node $j$ is the new patient zero as shown in Fig.~\ref{fig:vertices1}.
\end{example}

\subsection{Joint weights and sum rules}
We will derive the transition probability density $\W$ in Eq.~\eqref{eq:pathChain} from weights $J(\vs,\vg)$ defined in a joint space of paths and graphs. These \emph{joint weights} can again be expressed as products over local events,
\begin{equation}\label{eq:def-J}
	J(\vs,\vg)=C(\vs) \Big(\prod_{z\in\vg}\nu_z(g_z)\Delta(\vs_z,g_z) \Big)e^{-\int_0^T\ud t\,\Gamma^t}
\end{equation}
where $\nu_z(g_z)$ is a \emph{graph vertex rate}, which is multiplied by a \emph{compatibility indicator} $\Delta(\vs_z, g_z) \in \{0,1\}$ that enforces the compatibility of the vertex $g_z$ with the local state dynamics $\vs_z$ at location $z$. The local state dynamics $\vs_z$ can be one of the elementary state-changing events \eqref{eq:def-stateChange} or a \emph{null-event}
\begin{equation}\label{eq:def-nonEvent}
	\vs_z^{\const}:=(\vs_\vz^{t-}\to\vs_\vz^{t+})=(\vs_\vz^t\to \vs_\vz^t),
\end{equation}
meaning that the local state on nodes $\vz$ does not change at time $t$.
The exponential in Eq.~\eqref{eq:def-J} is a normalization factor, independent of $\vs$ and $\vg$, characterized by the \emph{total uniformization rate} $\Gamma^t$ which can also be decomposed into local rates $\Gamma_\vz^t\equiv \Gamma_z$ such that
\begin{equation}\label{eq:def-Gamma-total}
	\Gamma^t=\sum_\vz \Gamma_\vz^t,
\end{equation}
where the sum runs over all positions for which we have a graph vertex set \eqref{eq:def-graphVertices} at time $t$.

The graph vertex sets $\G_z$, vertex rates $\nu_z(g_z)$, and associated uniformization rates $\Gamma_z$ are to be chosen such that the trajectory probability density $P(\vs)$ is the marginalization of the joint weights $J(\vs,\vg)$ over all compatible graphs,
\begin{equation}\label{eq:Jmargin-g}
	P(\vs) \,= \sump_{\vg\ni\vs} J(\vs,\vg).
\end{equation}

The physical dimension of $P(\vs)$ is $(\text{time})^{-\Np_\vs}$ and the dimension of $J(\vs,\vg)$ is $(\text{time})^{-\Np_\vs-\Nb_{\vg,\vs}}$.
Correspondingly, the ``sum'' $\sum'_\vg$ over all graphs (compatible with $\vs$) is actually the path integral
\begin{align}\label{eq:def-graphSum}
	\sump_{\vg\ni\vs}&:=\sum_{g_{x_1},\dotsc,g_{x_\Np}} \sum_{\Nb=0}^\infty \frac{1}{\Nb!}\\ \nonumber
	&\times \int_0^T\!\!\! \ud t_1\dots\ud t_\Nb \!\!\!
	\sum_{\vy_1,\dotsc,\vy_\Nb}\,\sum_{g_{y_1}\in\G_{\vy_1}^{t_1}}\dots\!\!\sum_{g_{y_\Nb}\in\G_{\vy_\Nb}^{t_\Nb}}
\end{align}
where $x_1,\dotsc,x_\Np$ are the $\Np=\Np_\vs$ event locations for $\vs$ and the first sum runs over the associated vertex types $g_{x_1},\dotsc,g_{x_\Np}$. The sums in the second line run over all possible positions $\vy_m$ and vertex types $g_{y_m}$ of $\Nb=\Nb_{\vg,\vs}$ background graph vertices. Compatibility of $\vs$ with $\vg$ is implied on the right-hand side, which we could make explicit by including factors $\Delta(\vs_z, g_z)$, which are however already contained in the definition \eqref{eq:def-J} of $J(\vs,\vg)$.

To achieve Eq.~\eqref{eq:Jmargin-g}, we choose the local graph rates $\nu_z(g_z)$ such that they obey the local \emph{transition sum rule}
\begin{subequations}
\begin{equation}\label{eq:nu-transitionRule}
	\omega_z(\vs_z) = \sum_{g_z\in\G_z} \nu_z(g_z) \Delta(\vs_z, g_z)\ \ \forall\vs_z,
\end{equation}
and the local \emph{uniformization sum rule}
\begin{equation}\label{eq:nu-uniformSumRule}
	\Gamma_z - \Lambda_z(\vs_\vz^t) = \sum_{g_z\in\G_z} \nu_z(g_z)\Delta(\vs_z^{\const}, g_z)\ \ \forall\vs_\vz^t.
\end{equation}
\end{subequations}
In words, (a) each event rate must equal the rate sum of all compatible graph vertices, and (b) the rate sum of all (background) graph vertices that are compatible with the nodes remaining in their local state $\vs_\vz^t$ must perfectly fill the gap between the local escape rate $\Lambda_z(\vs_\vz^t)$ [Eq.~\eqref{eq:def-Lambda}] and the state-independent(!) local uniformization rate $\Gamma_z$.
Appendix~\ref{appx:Jmargin-g} proves the resulting marginalization condition \eqref{eq:Jmargin-g} for the joint weights $J(\vs,\vg)$.

\subsection{Transition probabilities and detailed balance}
For the two transitions $\vs\to\vg$ and $\vg\to\vs'$ that compose the cluster update \eqref{eq:clusterUpdate-sgs}, we use the joint weights \eqref{eq:def-J} to define the path-graph transition probability density 
\begin{subequations}\label{eq:def-W}
\begin{equation}\label{eq:def-Wgs}
	W(\vg|\vs) := \frac{J(\vs,\vg)}{P(\vs)}.
\end{equation}
and the graph-path transition probability
\begin{equation}\label{eq:def-Wsg}
	W(\vs'|\vg) := \frac{J(\vs',\vg)}{Q(\vg)},
\end{equation}
\end{subequations}
where the \emph{graph probability density} $Q(\vg)$ is the marginalization of $J(\vs,\vg)$ over all compatible states
\begin{equation}\label{eq:Jmargin-s}
	Q(\vg):=\sum_{\vs\in\vg}J(\vs,\vg).
\end{equation}

Note that, in contrast to the path integral $\sum'_{\vg\ni\vs}$ in the marginalization \eqref{eq:Jmargin-g} of $J$ over graphs, the sum $\sum_{\vs\in\vg}$ in Eq.~\eqref{eq:Jmargin-s} is really just a discrete sum: As previously discussed, a graph $\vg$ defines a finite set or, in the thermodynamic limit, at least a countable set of compatible trajectories. A constructed graph $\vg$ fully locks in the continuous degrees of freedom (the times of all vertices $z \in \vg$). The only remaining variables defining a compatible trajectories $\vs\in\vg$ are the discrete states assigned to the free clusters; see Sec.~\ref{sec:clusterIdentify}.

The choice of $W(\vg|\vs)$ and $W(\vs'|\vg)$ implies the detailed balance \eqref{eq:detail_balance} for the path-path transitions $\vs\to\vs'$ according to the path transition probability density
\begin{align}\nonumber
	\W(\vs'|\vs)&:=\sump_{\vg\ni\vs}W(\vs'|\vg)W(\vg|\vs)\\
	&\stackrel{\eqref{eq:def-W}}{=} \frac{1}{P(\vs)}\sump_{\vg\ni\vs}\frac{J(\vs',\vg)J(\vs,\vg)}{Q(\vg)}.
	\label{eq:def-Wss}
\end{align}
In particular, multiplying Eq.~\eqref{eq:def-Wss} by $P(\vs)$ and changing the path integral over $\vg\ni\vs$ to an integral over $\vg\ni\vs,\vs'$, in accordance with the fact that $J(\vs',\vg)=0$ if $\vg$ is incompatible with $\vs'$, we have
\begin{align}\nonumber
	\W(\vs'|\vs)P(\vs)=\sump_{\vg\ni\vs,\vs'}\frac{J(\vs',\vg)J(\vs,\vg)}{Q(\vg)}.
\end{align}
The right-hand side being manifestly symmetric in $\vs$ and $\vs'$, this establishes the detailed balance \eqref{eq:detail_balance}.

\subsection{Graph construction}\label{sec:grapConstruct}
Given a trajectory $\vs$, in step (a) of the cluster update, we want to construct a compatible graph $\vg$ according to the path-graph transition probability density $W(\vg|\vs)= J(\vs,\vg)/P(\vs)$ from Eq.~\eqref{eq:def-Wgs}. Recalling the discussion in Sec.~\ref{sec:clusterUpdateIntro}, we denote the list of $\vg$'s background vertices with respect to $\vs$ by $\vg^\bg_\vs$ and their locations by $y\in\vg^\bg_\vs$ (all vertices in $\vg$, where there is no state change in $\vs$).

\Emph{Transition probability.}~-- Plugging in the expressions \eqref{eq:def-P} and \eqref{eq:def-J} for the path probability density $P(\vs)$ and the joint weights $J(\vs,\vg)$, we get
\begin{equation}\label{eq:Wgs}
	W(\vg|\vs)= W_\bg(\vg^\bg_\vs|\vs)\, \prod_{x\in\vs}\frac{\nu_x(g_x)\Delta(\vs_x, g_x)}{\omega_x(\vs_x)}
\end{equation}
with the \emph{background density}
\begin{equation*}
	W_\bg(\vg^\bg_\vs|\vs) =\Big(\!\prod_{y\in\vg^\bg_\vs} \nu_y(g_y)\Delta(\vs_y,g_y)\Big)
	 e^{\int_0^T\ud t\,\big[\Lambda^t(\vs^t)-\Gamma^t\big]}.
\end{equation*}
For the exponential factor, we can decompose the total escape and uniformization rates into local rates $\Lambda^t(\vs^t)=\sum_\vy \Lambda_y(\vs_\vy^t)$ and $\Gamma^t=\sum_\vy \Gamma_y$ according to Eqs.~\eqref{eq:def-Lambda-total} and \eqref{eq:def-Gamma-total} and plug in the uniformization sum rule \eqref{eq:nu-uniformSumRule}. For the compatibility factors, we can replace $\vs_y$ by the null-events $\vs_y^{\const}$ [Eq.~\eqref{eq:def-nonEvent}] because background events occur exclusively where the state is constant. In this way, we find that the background density is the measure for a multi-dimensional inhomogeneous Poisson point process \cite{Daley2003} with time-and-position-dependent rates $\lambda(y,g_y)=\nu_y(g_y)\Delta(\vs_y^{\const},g_y)$,
\begin{align}\label{eq:Wgs-bg}
	W_\bg&(\vg^\bg_\vs|\vs) =\Big(\!\prod_{y\in\vg^\bg_\vs} \nu_y(g_y)\Delta(\vs_y^{\const},g_y)\Big) \\ \nonumber
	&\!\!\times \exp\Big(-\int_0^T \!\!\ud t \sum_\vy\sum_{g'_y\in\G_\vy^t} \nu_y(g'_y)\Delta(\vs_y^{\const},g'_y)\Big).
\end{align}

\Emph{Graph generation.}~-- The results \eqref{eq:Wgs} and \eqref{eq:Wgs-bg} correspond to a construction of a compatible graph $\vg$ in two phases:
\begin{enumerate}
 \item[(i)]
 According to the product term in the path-graph transition probability density \eqref{eq:Wgs} and employing the transition sum rule \eqref{eq:nu-transitionRule} for the denominators $\omega_x(\vs_x)$, we loop over all event locations $x$ in $\vs$ and assign graph vertex $g_x$ with probability
 \begin{equation}
	W_x(g_x|\vs_x) := \frac{\nu_x(g_x)\Delta(\vs_x,g_x)}{\sum_{g'_x\in\G_x} \nu_x(g'_x)\Delta(\vs_x,g'_x)}.
 \end{equation}
 \item[(ii)]
 For all other locations $y$ where no event occurs in $\vs$, background graph vertices are inserted through an inhomogeneous Poisson point process. Specifically, for each possible type of vertex $g$, we identify all space-time intervals where it would be compatible, and randomly add instances with temporal density $\nu_y(g)$ to $\vg$. If the rate $\nu_y(g)$ is uniform on a time interval $[T_a,T_b]$, we can select the number $\Nb$ of type-$g$ background vertices according to the discrete Poisson distribution $\mu^{\Nb}e^{-\mu}/\Nb!$ with $\mu=(T_b-T_a)\nu_y(g)$ and, then, place vertex $g$ at $\Nb$ times $t_1,\dotsc,t_\Nb$ on the interval according to the uniform random distribution.
\end{enumerate}

\subsection{Cluster identification and state update}\label{sec:clusterIdentify}
In step (b) of the cluster update \eqref{eq:clusterUpdate-sgs}, we want to determine a new trajectory $\vs'$ that is compatible with the generated graph $\vg$ according to the graph-path transition probability $W(\vs'|\vg)= J(\vs',\vg)/\big(\sum_{\vs\in\vg}J(\vs,\vg)\big)$ from Eq.~\eqref{eq:def-Wsg}. 

\Emph{Transition probability.}~-- Writing the joint weight \eqref{eq:def-J} in the form
\begin{align}
	J(\vs',\vg) &\,= C(\vs')D(\vs',\vg) K(\vg)\quad\text{with}\\
	D(\vs',\vg) &:= \prod_{z\in\vg}\Delta(\vs'_z,g_z)
\end{align}
being the \emph{compatibility indicator} and $K(\vg)$ collecting all $\vs'$ independent terms, the transition probability assumes the simple form
\begin{equation}\label{eq:Wsg}
	W(\vs'|\vg)= \frac{C(\vs')D(\vs',\vg)}{\sum_{\vs\in\vg} C(\vs)D(\vs,\vg)}.
\end{equation}

The result \eqref{eq:Wsg} means that we should select among all trajectories $\{\vs'\in\vg\}$ that are compatible with the graph $\vg$ according to the probability distribution $C(\vs')$. Recall that $C$ can be an indicator function for trajectory constraints like having a big epidemic outbreak or, more generally, a probability distribution to favor trajectories of a certain type during the sampling. In this section, we assume the trivial case, where $C(\vs')=1$ for all $\vg$-compatible trajectories such that we should select among them with uniform probability. Modifications due to nontrivial trajectory constraints are discussed in Sec.~\ref{sec:BC}.

\Emph{Cluster identification.}~-- 
To efficiently translate the graph $\vg$ into spacetime clusters $\C_k$ and determine their allowed states
\begin{equation}
	a_k\in\A(R_k)\subset\{0,\dotsc,d-1\}
\end{equation}
we can employ the union-find (a.k.a.\ disjoint-set) algorithm \cite{Galler1964-7,Tarjan1975-22,Tarjan1984-31,Brass2008}. Because the state of any node is strictly constant between the graph vertices, we can divide the continuous-time line of each node into discrete segments and use the union-find algorithm to group them according to the rules of $\vg$:
\begin{itemize}
 \item 
 For every node $i\in\{1,\dotsc,N\}$, collect the times $t_1 < t_2 < \dots < t_{M-1}$ of all graph vertices in $\vg$ where node $i$ is involved. In conjunction with $t_0:=0$ and $t_M:=T$, these times decompose the trajectory of node $i$ into a sequence of segments $(i, t_m, t_{m+1})$ on which the state $\sigma^{\prime t}_i$ is constant.
 \item
 We then create a union-find data structure, where every segment is initialized as its own independent root $R$ (its own cluster with $\textsc{Parent}(R)\leftarrow R$). For every root $R$, we initialize the set of allowed states as $\A(R)=\{0,1,\dotsc,d-1\}$ and assign the trivial identity map from cluster to node states ($\textsc{Map}(R)\leftarrow \operatorname{Id}$).
 \item
 Now, iterate through the graph vertices $g_z$ at locations $z=(t,\vz)\in\vg$ to bind the segments together and apply constraints. A vertex $g_z$ can impose bijective relations among some of the participating variables $\vec{s}=(\vs_\vz^{t-},\vs_\vz^{t+})$ (e.g., imposing that two variables are equal) and restrict the sets of valid states for these variables as detailed in Sec.~\ref{sec:SIS-vertexSets} for SIS models. A bijective relation $s_k=f(s_\ell)$ between two variables $s_k$ and $s_\ell$ corresponds to a cluster unification. Unless the two variables were already part of the same cluster, we attach one cluster tree to the other by making one of the roots, say $R_k$ the parent of the other ($\textsc{Parent}(R_\ell)\leftarrow R_k$), store the associated bijective relation ($\textsc{Map}(R_\ell)\leftarrow f$) and reduce the available state set to the intersection
 \begin{equation}
	\A(R_k)\leftarrow \A(R_k)\cap f\big(\A(R_\ell)\big).
 \end{equation}
 If the graph vertex explicitly restricts the allowed states for a variable $s_k$ to $\A_{g_z,k}$, we restrict the allowed state set of the cluster accordingly
 \begin{equation}
	\A(R_k) \leftarrow \A(R_k) \cap \A_{g_z,k}.
 \end{equation}
\end{itemize}

\Emph{Cluster state update.}~--
The clusters $\C_k$, identified by the tree roots with $\textsc{Parent}(R_k)=R_k$, decompose the full spacetime volume $[0,T]\times\{1,\dotsc,N\}$ into connected components. For the transition $\vg \to \vs'$ to the new trajectory $\vs'$, we identify all ``free'' clusters with $d_k\equiv|\A(R_k)|>1$. In Figs.~\ref{fig:vertices1}-\ref{fig:E2update}, these are indicated by red lines. The states on each free cluster $\C_k$ are then updated by either retaining the original state from $\vs$ or choosing one of the other $d_k-1$ allowed states with equal probability $1/d_k$.
This cluster-flip naturally generates an updated valid trajectory $\vs'$ without requiring subsequent time-step re-evaluations, bypassing critical slowing down.

\section{Graph vertex sets for  SIS models}\label{sec:SIS-vertexSets}
In the following, let us consider SIS models, where recovery rates $\gamma_i$ and $\alpha_{i,j}$ are allowed to be position-dependent and asymmetric ($\alpha_{i,j}\neq\alpha_{j,i}$). For simplicity, we do not denote any time dependence of these rates.

Our goal is to find sets of single-node graph vertices $\G_{(t,i)}=\{g_{(t,i)}\}$ and edge graph vertices $\G_{(t,i,j)}=\{g_{(t,i,j)}\}$, associated vertex rates $\nu_x(g_x)$, and uniformization rates $\Gamma_x$ such that the transition and uniformization sum rules \eqref{eq:nu-transitionRule} and \eqref{eq:nu-uniformSumRule} are obeyed, and we obtain an ergodic algorithm in the sense that the cluster updates allow us to reach any physically valid trajectory $\vs$. According to our convention from Sec.~\ref{sec:pathProb} we have $i<j$.

\subsection{Single-node vertex set \textsc{S1}}
For the single-node vertices, related to recovery events, we can choose
\begin{equation}
	\G_x=\left\{g^{0}_x,\,g^{\tR+}_x,\,g^{\tR-}_x\right\}
\end{equation}
with $x=(t,i)$ and
\begin{subequations}\label{eq:SIS-S1-G}
\begin{alignat}{3}
	&g^{0}_x   &&=(0\to 0),\\
	&g^{\tR+}_x&&=(a\to 0),\\
	&g^{\tR-}_x&&=(1\to a),
\end{alignat}
\end{subequations}
where $a\in\{0,1\}$ indicates that the corresponding node-$i$ state $\sigma_i^{t-}$ just before the vertex time $t$ (for $g^{\tR+}_x$) or $\sigma_i^{t+}$ just after time $t$ (for $g^{\tR-}_x$) remains unconstrained by the vertex. The vertices $g^{\tR\pm}_x$ were already discussed in Example~\ref{ex:vertices1}. The vertex $g^{0}_x$ is only compatible with the null-event, where node $i$ stays susceptible, i.e., $\sigma_i^{t-}=\sigma_i^{t+}=0$.

Denoting the associated vertex rates by $\nu^{0}_x$ and $\nu^{\tR\pm}_x$, the transition sum rule \eqref{eq:nu-transitionRule} for the only physically possible state change $(1\to 0)$ reads
\begin{equation}
	\gamma_i=\omega_x(1\to 0) = \nu^{\tR+}_x + \nu^{\tR-}_x,
\end{equation}
and, with the local escape rates $\Lambda_x(0)=0$ and $\Lambda_x(1)=\gamma_i$, the uniformization sum rules \eqref{eq:nu-uniformSumRule} for states $\sigma_i^t=0$ and $\sigma_i^t=1$ impose the constraints
\begin{equation}
	\Gamma_x=\nu^{0}_x + \nu^{\tR+}_x,\quad 
	\Gamma_x-\gamma_i=\nu^{\tR-}_x.
\end{equation}

This is a system of three linear equations for four parameters. Choosing $\varphi:=\nu^{0}_x$ as the free parameter, the solution space is
\begin{equation}\label{eq:phi}
	\left(\Gamma_x,\,\nu^{0}_x,\,\nu^{\tR+}_x,\,\nu^{\tR-}_x\right)
	=\left(\gamma_i+\frac{\varphi}{2},\,\varphi,\,\gamma_i-\frac{\varphi}{2},\,\frac{\varphi}{2}\right)
\end{equation}
with $0\leq\varphi\leq 2\gamma_i$ as all rates need to be nonnegative.

In order to minimize the computation time per cluster update, it is generally best to minimize the uniformization rate $\Gamma_x$ such that, in light of the uniformization sum rule \eqref{eq:nu-uniformSumRule} and the background density \eqref{eq:Wgs-bg}, the total background-vertex insertion rates for every state are as small as possible. This leads to the optimum at $\varphi=0$ with
\begin{equation}\label{eq:phi0}
	\left(\Gamma_x,\,\nu^{0}_x,\,\nu^{\tR+}_x,\,\nu^{\tR-}_x\right)
	=\left(\gamma_i,\,0,\,\gamma_i,\,0\right),
\end{equation}
implying that we would actually work with only one single-node vertex, $\G_x=\{g^{\tR+}_x\}$ which can prolong infections or introduce new infections. The infection-shortening vertex $g^{\tR-}_x$ would be gone, implying that infection intervals can only be removed in their entirety or modified by edge vertices.
Concerning numerical efficiency, we also need to consider auto-correlations in the Markov chain \eqref{eq:pathChain} of trajectories. While the $\varphi=0$ solution \eqref{eq:phi0} generally has the lowest computation cost per update, it may imply longer auto-correlation times or, depending on the constraint $C(\vs)$ and its implementation, even fail ergodicity.

Alternatively, we could choose the maximum $\varphi=2\gamma_i$ such that vertex $g^{\tR+}_x$ is removed. However, this would imply that we cannot extend infection times or increase the number of infected nodes at $t=0$.

In our experience, the solution \eqref{eq:phi} with $\varphi=\gamma_i/2$ is a solid choice.

\subsection{Edge vertex set \textsc{E1}}\label{sec:SIS-E1}
For the vertices on edges $(i,j)$, related to infection events, we can choose
\begin{equation}
	\G_x=\left\{g^\taa_x,\,g^{\ta\up-}_x,\,g^{\ta\dw-}_x,\,g^{\tI\up-}_x,\,g^{\tI\dw-}_x,\,g^{\tI\up+}_x,\,g^{\tI\dw+}_x\right\}
\end{equation}
with $x=(t,i,j)$ and
\begin{subequations}\label{eq:SIS-E1-G}
\begin{alignat}{3}
	&g^\taa_x&&=(a,a\to a,a),\\
	&g^{\ta\up-}_x	&&=(a,0\to a,a),\\
	&g^{\ta\dw-}_x	&&=(0,a\to a,a),\\
	&g^{\tI\up-}_x	&&=(1,0\to 1,a),\\
	&g^{\tI\dw-}_x	&&=(0,1\to a,1),\\
	&g^{\tI\up+}_x	&&=(1,a\to 1,1),\\
	&g^{\tI\dw+}_x	&&=(a,1\to 1,1).
\end{alignat}
\end{subequations}
Again $a\in\{0,1\}$ indicates that the corresponding node state just before or just after the vertex time $t$ remains unconstrained by the vertex, but if variable $a$ occurs multiple times in a vertex, it imposes equality among the corresponding node states. In particular, the vertex $g^\taa_x$ matches the two null-events where the edge stays either in the $(0,0)$ or in the $(1,1)$ state. The vertex $g^{\ta\up-}_x$ matches the null-event $(0,0\to 0,0)$ and the infection event $(1,0\to 1,1)$.

Denoting the associated vertex rates by $\nu^\taa_x$, $\nu^{\ta\up-}_x$, $\nu^{\ta\dw-}_x$, $\nu^{\tI\up-}_x$, $\nu^{\tI\dw-}_x$, $\nu^{\tI\up+}_x$, and $\nu^{\tI\dw+}_x$, the transition sum rules \eqref{eq:nu-transitionRule} for the two infection-related state changes $(1,0\to 1,1)$ and $(0,1\to 1,1)$
read
\begin{subequations}\label{eq:SIS-E1-trans}
\begin{align}
	\alpha_{i,j}&=\omega_x(1,0\to 1,1) = \nu^{\ta\up-}_x + \nu^{\tI\up-}_x + \nu^{\tI\up+}_x,\\
	\alpha_{j,i}&=\omega_x(0,1\to 1,1) = \nu^{\ta\dw-}_x + \nu^{\tI\dw-}_x + \nu^{\tI\dw+}_x,
\end{align}
\end{subequations}
and, with the local escape rates 
\begin{subequations}\label{eq:SIS-E-Lambda}
\begin{gather}
	\Lambda_x(0,0)=\Lambda_x(1,1)=0,\\
	\Lambda_x(1,0)=\alpha_{i,j},\quad\text{and}\quad
	\Lambda_x(0,1)=\alpha_{j,i},
\end{gather}
\end{subequations}
the uniformization sum rules \eqref{eq:nu-uniformSumRule} for the states $(0,0)$, $(1,1)$, $(1,0)$, and $(0,1)$ impose the constraints
\begin{subequations}\label{eq:SIS-E1-uni}
\begin{gather}
	\Gamma_x=\nu^\taa_x + \nu^{\ta\up-}_x + \nu^{\ta\dw-}_x,\\
	\Gamma_x=\nu^\taa_x + \nu^{\tI\up+}_x + \nu^{\tI\dw+}_x,\\
	\Gamma_x-\alpha_{i,j}=\nu^{\tI\up-}_x,\\
	\Gamma_x-\alpha_{j,i}=\nu^{\tI\dw-}_x.
\end{gather}
\end{subequations}

Equations~\eqref{eq:SIS-E1-trans} and \eqref{eq:SIS-E1-uni} form a system of six linear equations for eight parameters. Choosing $\vartheta^\up:=\nu^{\ta\up-}_x$ and $\vartheta^\dw:=\nu^{\ta\dw-}_x$ as the free parameters, the solution space is
\begin{align}\nonumber
	&\left(\Gamma_x,\,\nu^\taa_x,\,\nu^{\ta\up-}_x,\,\nu^{\ta\dw-}_x,\,\nu^{\tI\up-}_x,\,\nu^{\tI\dw-}_x,\,\nu^{\tI\up+}_x,\,\nu^{\tI\dw+}_x\right)\\
	&\textstyle \nonumber
	=\big(2\bar{\alpha}-2\bar{\vartheta},\, 2\bar{\alpha}-4\bar{\vartheta},\, \vartheta^\up,\, \vartheta^\dw,\\
	&\qquad \alpha_{j,i}-2\bar{\vartheta},\, \alpha_{i,j}-2\bar{\vartheta},\, \vartheta^\dw+\Delta\alpha,\, \vartheta^\up-\Delta\alpha \big)
	\label{eq:SIS-E1-sol}
\end{align}
with
\begin{equation}\label{eq:alphaRel}\textstyle
	\bar{\alpha}:=\frac{\alpha_{i,j}+\alpha_{j,i}}{2},\quad
	\Delta\alpha:=\alpha_{i,j}-\alpha_{j,i},\quad
	\bar{\vartheta}:=\frac{\vartheta^\up+\vartheta^\dw}{2}
\end{equation}
and $\bar{\vartheta}$ limited to the range
\begin{equation}\label{eq:SIS-E1-theta}
	|\Delta\alpha| \leq 2\bar{\vartheta} \leq \min\left(\alpha_{i,j},\alpha_{j,i}\right),
\end{equation}
which follows from the constraint that all rates need to be nonnegative.
\begin{figure*}[t]
\label{fig:E1upLocks}
\centering
\includegraphics[width=0.95\textwidth]{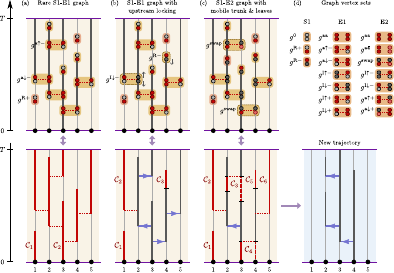}
\caption{\textbf{Upstream lock avalanches with edge vertex set \textsc{E1} and unlocking with set \textsc{E2}.} For a given SIS-model trajectory $\vs$, panels (a), (b), and (c) show different compatible graphs $\vg$ (top) and the associated configurations of clusters (bottom). We employ the same graphical conventions as in Fig.~\ref{fig:vertices1}, and panel (d) specifies the different employed vertex sets in accordance with Sec.~\ref{sec:SIS-vertexSets}.
Panel (a) shows an (unlikely) choice of graph vertices that lead to a big free cluster $\C_2$ that includes patient zero (node 3). To achieve a mobile patient zero with vertex sets \textsc{S1} and \textsc{E1}, the graph must happen to be such that all vertices downstream of patient zero are $g^{\tR+}$, $g^{\ta\up-}$, or $g^{\ta\dw-}$ as in this case. The probability of this decreases exponentially in the length of the infection tree. 
As illustrated in panel (b), a single downstream $g^{\tI\up-}$, $g^{\tI\dw-}$ or $g^{\tR-}$ leads to a cascade of locks that fix patient zero. Such upstream lock avalanches lead to a general inflexibility of the infection tree's ``trunk'' and extended autocorrelations in the path Markov chain \eqref{eq:pathChain}.
As exemplified in panel (c) and discussed in Sec.~\ref{sec:SIS-E2}, the alternative edge vertex set \textsc{E2} resolves this issue. Flipping clusters $\C_2,\dotsc,\C_6$ in the shown example leads to a very different trajectory $\vs'$, where node 4 is the new patient zero.
Note that, for clarity, we have not used all available graph vertices here or tried to show particularly realistic examples.}
\end{figure*} 

The admissible range \eqref{eq:SIS-E1-theta} is actually only nonempty if $|\Delta\alpha| \leq \min(\alpha_{i,j},\alpha_{j,i})$, i.e., only for moderately asymmetric infection rates with
\begin{equation}\label{eq:SIS-E1-asymRestrict}
	\frac{1}{2} \leq \frac{\alpha_{i,j}}{\alpha_{j,i}} \leq 2
	\quad\text{or}\quad
	\frac{1}{2} \leq \frac{\alpha_{j,i}}{\alpha_{i,j}} \leq 2.
\end{equation}
To lift this restriction, we will consider alternative edge-vertex sets in Secs.~\ref{sec:SIS-E2} and \ref{sec:SIS-E3}.

For the rest of this section, let us assume symmetric infection and vertex rates, i.e.,
\begin{equation}
	\alpha_{i,j}=\alpha_{j,i}=:\alpha,\quad
	\vartheta^\up=\vartheta^\dw=:\vartheta.
\end{equation}
How should we choose $\vartheta\in[0,\alpha/2]$ to make CPMC as efficient as possible? The answer is not obvious.

The uniformization rate $\Gamma_x=2\alpha-2\vartheta$ and, hence, the expected computation time per cluster update are minimized for $\vartheta=\alpha/2$ with
\begin{align}\nonumber
	&\left(\Gamma_x,\,\nu^\taa_x,\,\nu^{\ta\up-}_x,\,\nu^{\ta\dw-}_x,\,\nu^{\tI\up-}_x,\,\nu^{\tI\dw-}_x,\,\nu^{\tI\up+}_x,\,\nu^{\tI\dw+}_x\right)\\
	& =\left(\alpha, 0,\, \frac{\alpha}{2},\, \frac{\alpha}{2},\, 0,\, 0,\, \frac{\alpha}{2},\, \frac{\alpha}{2} \right),
	\label{eq:SIS-E1-thetaMax}
\end{align}
removing the vertices $g^\taa_x$, $\nu^{\tI\up-}_x$, and $\nu^{\tI\dw-}_x$. Hence, we are left with the reduced vertex set $\G_x=\left\{g^{\ta\up-}_x,g^{\ta\dw-}_x,g^{\tI\up+}_x,g^{\tI\dw+}_x\right\}$.

The other extreme $\vartheta=0$ maximizes the uniformization rate with
\begin{align}\nonumber
	&\left(\Gamma_x,\,\nu^\taa_x,\,\nu^{\ta\up-}_x,\,\nu^{\ta\dw-}_x,\,\nu^{\tI\up-}_x,\,\nu^{\tI\dw-}_x,\,\nu^{\tI\up+}_x,\,\nu^{\tI\dw+}_x\right)\\
	& =\left(2\alpha,\, 2\alpha,\, 0,\, 0,\, \alpha,\, \alpha,\, 0,\, 0 \right)
	\label{eq:SIS-E1-theta0}
\end{align}
such that we would work with the reduced vertex set $\G_x=\left\{g^\taa_x,g^{\tI\up-}_x,g^{\tI\dw-}_x\right\}$. Note that vertices $g^{\tI\up-}_x$ and $g^{\tI\dw-}_x$ leave the future state of an infected node unconstrained, which may be computationally beneficial for the simulation of big outbreaks.

\subsection{Drawbacks of edge vertex set \textsc{E1}}\label{sec:SIS-E1-analysis}
The edge vertex set \textsc{E1} has two important drawbacks.

\Emph{Restriction on asymmetry}:
It leads to the restriction \eqref{eq:SIS-E1-asymRestrict} on the asymmetry of infection rates $\alpha_{i,j}$ and $\alpha_{j,i}$, making the vertex set applicable only if the two rates vary by less than a factor of two relative to each other.

\Emph{Constraining the epidemic trunk}:
In a scenario, where we start with a single infected node at time $t=0$ (patient zero) which infects further nodes, the cluster updates with vertex sets \textsc{S1} and \textsc{E1} alone would make updates of the patient-zero node very unlikely. While $g^{\ta\up-}_x=(a,0\to a,a)$ and $g^{\ta\dw-}_x$ vertices in principle allow patient zero to be in a free cluster, this will only happen under the condition that all downstream infections in the infection tree get assigned this type of graph vertex instead of $g^{\tI\up-}_x=(1,0\to 1,a)$, $g^{\tI\up+}_x=(1,a\to 1,1)$, or their $\dw$ counterparts. But the latter vertices will be selected with a finite probability and background single-node vertices $g^{\tR-}_y=(1\to a)$ will be dropped by the Poisson process, locking in the infected state upstream. The single-node vertex $g^{\tR+}_x=(a\to 0)$ cannot help either. Hence, the probability of a patient zero residing in a free cluster decays exponentially in the ``length'' of the infection tree (the number of downstream infection events). This is illustrated in panels (a) and (b) of Fig.~\ref{fig:E1upLocks}.
The issue extends beyond patient zero -- it heavily constrains the entire ``trunk'' of the infection tree due to such upstream \emph{lock avalanches}, restricting the cluster updates to beautifully fluctuate the surface of an epidemic tree but making it difficult to mutate its skeleton. While this can be substantially alleviated by using special update rules for patient zero (see Sec.~\ref{sec:BC}) and employing parallel tempering \footnote{J.\ Sun and T.\ Barthel, in preparation}, it is more attractive to resolve the issue by an adaptation of the edge vertex set.

\subsection{Edge vertex set \textsc{E2} -- unlocking the epidemic trunk}\label{sec:SIS-E2}
Let us first assess our options to alleviate the drawbacks of vertex set \textsc{E1} described in Sec.~\ref{sec:SIS-E1-analysis}. As detailed in Appendix~\ref{appx:SIS-completeEdgeSet} the vertices $g^{\ta\up-}_x$, $g^{\ta\dw-}_x$, $g^{\tI\up-}_x$, $g^{\tI\dw-}_x$, $g^{\tI\up+}_x$, $g^{\tI\dw+}_x$ from Eq.~\eqref{eq:SIS-E1-G} in combination with the new vertices
\begin{subequations}\label{eq:SIS-E-addedVertices}
\begin{alignat}{3}
	&g^\swap_x		&&=(a,\bar{a}\to 1,1),\\
	&g^{\ta\up+}_x	&&=(a,\bar{a}\to a,1),\\
	&g^{\ta\dw+}_x	&&=(\bar{a},a\to 1,a)
\end{alignat}
represent all nine options for vertices that are compatible with an infection event and feature at least one free variable $a\in\{0,1\}$ with $\bar{a}$ denoting the negation of $a$.
Let us also add the null-event vertex
\begin{equation}
	g^{\ta\bta}_x=(a,\bar{a}\to a,\bar{a})
\end{equation}
\end{subequations}
to the list.

To achieve a large number of free clusters:
\begin{itemize}
 \item 
 Let us avoid all vertices that fix all four of the involved $t^-$ and $t^+$ variables, like $(1,0\to 1,1)$.
 \item
 The swap vertex $g^\swap_x=(a,\bar{a}\to 1,1)$ prevents upstream lock avalanches through temporal decoupling, allowing patient zero to move and, more generally, infection sources to swap. The input variables $(a,\bar{a})$ are completely disconnected from the outputs $(1,1)$. If a subsequent event locks the future of node $i$ or $j$ to $1$, that lock hits the $(1,1)$ output of this vertex and cannot cross the event boundary to lock the past.
 \item
 The edge vertices $g^{\tI\up-}_x=(1,0\to 1,a)$ and $g^{\tI\dw-}_x$ in combination with the single-node vertices $g^{\tR-}_y=(1\to a)$ prevent downstream lock avalanches. While the swap vertex locks the immediate downstream future of the edge's nodes to 1 (infected), background recovery vertices $g^{\tR-}_y$ can convert locked nodes back to a free variable $a$ for the remainder of the timeline; $g^{\tI\up-}_x$ and $g^{\tI\dw-}_x$ can generate free target nodes downstream and allow for the conversion between infection events and null-events.
 \item
 Let us avoid the vertices $g^{\ta\up-}_x=(a,0\to a,a)$ and $g^{\ta\dw-}_x$, because they impose equality for three of the involved states and fix the fourth, which can lead to a propagation of locks in both time directions.
 \item 
 While the vertices $g^{\ta\up+}_x=(a,\bar{a}\to a,1)$ and $g^{\ta\dw+}_x$ are similar in this respect, we keep them as they will be useful for SIS models with asymmetric infection rates $\alpha_{i,j}\neq\alpha_{j,i}$. They are also useful because, in contrast to $g^\swap_x$, $g^{\tI\up-}_x$, and $g^{\tI\dw-}_x$, they do not fix the future of the source node and allow for uninterrupted flippable time segments of an infection source node.
\end{itemize}

\begin{figure*}[t]
\label{fig:E2update}
\centering
\includegraphics[width=\textwidth]{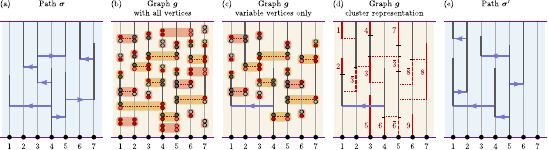}
\caption{\textbf{Example for a more complex cluster update with edge vertex set \textsc{E2}.} (a) An SIS-model trajectory $\vs$ on seven nodes with node 4 as patient zero and four infected nodes at the final time. (b) A compatible graph $\vg$ based on the \textsc{S1} and \textsc{E2} vertex sets.  As before, vertices at state-changing events in $\vs$ are indicated by orange boxes and background vertices are indicated by red boxes. (c) To make the cluster identification easier for the human eye, we have removed all vertices that ultimately have no free state variables and have fixed the corresponding parts of the trajectory. (d) Cluster representation of the graph $\vg$, where each red solid time segment with label ``$k$'' carries state $a_k=0,1$ of a free cluster $\C_k$ and each red dashed time segment with label ``$\bar{k}$'' carries the negated state $\bar{a}_k=1,0$ of cluster $\C_k$. (e) Flipping the states of clusters $\C_1$, $\C_2$, $\C_3$, $\C_4$, $\C_6$, $\C_7$, and $\C_8$ (with respect to their state in $\vs$), we arrive at the new trajectory $\vs'$, where both the trunk and leaves of the infection tree have changed substantially. The new patient zero is node 5, and there are now three infected nodes at the final time.}
\end{figure*}

Including the additional null-event vertex $g^{\ta\bta}_x$, these considerations lead us to the edge vertex set \textsc{E2}
\begin{equation}
	\G_x=\left\{g^\taa_x,\, g^{\ta\bta}_x,\, g^\swap_x,\, g^{\tI\up-}_x,\, g^{\tI\dw-}_x,\, g^{\ta\up+}_x,\, g^{\ta\dw+}_x\right\}
\end{equation}
with $x=(t,i,j)$. The resulting trajectory flexibility in cluster updates is illustrated in Figs.~\ref{fig:E1upLocks}c and \ref{fig:E2update}.

Denoting the associated vertex rates by $\nu^\taa_x$, $\nu^{\ta\bta}_x$, $\nu^\swap_x$, $\nu^{\tI\up-}_x$, $\nu^{\tI\dw-}_x$, $\nu^{\ta\up+}_x$, and $\nu^{\ta\dw+}_x$, the transition sum rules \eqref{eq:nu-transitionRule} for the two infection-related state changes $(1,0\to 1,1)$ and $(0,1\to 1,1)$
read
\begin{subequations}\label{eq:SIS-E2-trans}
\begin{align}
	\alpha_{i,j}&=\omega_x(1,0\to 1,1) = \nu^\swap_x + \nu^{\ta\up+}_x + \nu^{\tI\up-}_x ,\\
	\alpha_{j,i}&=\omega_x(0,1\to 1,1) = \nu^\swap_x + \nu^{\ta\dw+}_x + \nu^{\tI\dw-}_x.
\end{align}
\end{subequations}
With the local escape rates \eqref{eq:SIS-E-Lambda}, the uniformization sum rules \eqref{eq:nu-uniformSumRule} for the states $(0,0)$, $(1,1)$, $(1,0)$, and $(0,1)$ impose the constraints
\begin{subequations}\label{eq:SIS-E2-uni}
\begin{gather}
	\Gamma_x=\nu^\taa_x, \label{eq:SIS-E2-uni00}\\
	\Gamma_x=\nu^\taa_x, \label{eq:SIS-E2-uni11}\\
	\Gamma_x-\alpha_{i,j}=\nu^{\ta\bta}_x+\nu^{\ta\dw+}_x+\nu^{\tI\up-}_x,\\
	\Gamma_x-\alpha_{j,i}=\nu^{\ta\bta}_x+\nu^{\ta\up+}_x+\nu^{\tI\dw-}_x.
\end{gather}
\end{subequations}

As Eqs.~\eqref{eq:SIS-E2-uni00} and \eqref{eq:SIS-E2-uni11} are identical, we have a system of five linear equations for eight parameters. Choosing the uniformization rate $\Gamma_x$, $\psi:=\nu^\swap_x$, and $\phi:=\nu^{\tI\up-}_x$ as the free parameters, the solution space is
\begin{align}\nonumber
	&\left(\Gamma_x,\, \nu^\taa_x,\, \nu^{\ta\bta}_x,\, \nu^\swap_x,\, \nu^{\tI\up-}_x,\, \nu^{\tI\dw-}_x,\, \nu^{\ta\up+}_x,\, \nu^{\ta\dw+}_x\right)\\
	&\textstyle \nonumber
	=\Big(\Gamma_x,\, \Gamma_x,\, \Gamma_x-\alpha_{i,j}-\alpha_{j,i} + \psi,\, \psi, \phi, \phi,\\
	& \qquad \alpha_{i,j}-\psi-\phi, \alpha_{j,i}-\psi-\phi \Big)
	\label{eq:SIS-E2-sol}
\end{align}
with
\begin{subequations}\label{eq:SIS-E2-solRange}
\begin{gather}
	\Gamma_x \geq \alpha_{i,j} + \alpha_{j,i} - \psi\quad\text{and}\\
	0 \leq \psi + \phi \leq \min(\alpha_{i,j}, \alpha_{j,i})
\end{gather}
\end{subequations}
following from the constraint that all rates need to be nonnegative.

It is generally desirable to reduce the uniformization rate $\Gamma_x$. Doing so removes the $g^{\ta\bta}_x$ vertex and leaves us with the two free parameters $\psi$ and $\phi$. With $\psi=\nu^\swap_x$ primarily controlling upstream flexibility and $\phi=\nu^{\tI\up-}_x=\nu^{\tI\dw-}_x$ controlling downstream flexibility, a pragmatic choice could be to choose some $0\leq\psi\leq\min(\alpha_{i,j}, \alpha_{j,i})$ and set $\phi=\min(\alpha_{i,j}, \alpha_{j,i})-\psi$, with the $\psi$-optimum depending on the network structure.

In contrast to the restriction \eqref{eq:SIS-E1-asymRestrict} of vertex set \textsc{E1}, Eq.~\eqref{eq:SIS-E2-solRange} shows that set \textsc{E2} works for any ratio of the infection rates $\alpha_{i,j}$ and $\alpha_{j,i}$. However, on edges with extreme asymmetry where one of the infection rates is zero and the other nonzero, we have to choose $\psi=\phi=0$. This removes the swap vertex $g^\swap_x=(a,\bar{a}\to 1,1)$, which is natural as we cannot swap infection sources on such an edge. The removal of the $g^{\tI\up-}_x$ and $g^{\tI\dw-}_x$ vertices is however problematic as they are an important antidote against downstream lock avalanches and are needed to prune leafs of the infection tree.

\subsection{Edge vertex set \textsc{E3} -- unlocking full asymmetry}\label{sec:SIS-E3}
The problems of vertex set \textsc{E2} for strongly asymmetric infection rates $\alpha_{i,j}\neq\alpha_{j,i}$ can be resolved by introducing the asymmetry-nullifying (blocker) vertices
\begin{subequations}
\begin{alignat}{3}
	&g^{10}_x	&&=(1,0\to 1,0),\\
	&g^{01}_x	&&=(0,1\to 0,1).
\end{alignat}
\end{subequations}
Also removing the null-event vertex $g^{\ta\bta}_x$, which turned out as inessential for the set \textsc{E2}, we obtain the modified edge vertex set
\begin{equation}
	\G_x=\left\{g^\taa_x,\,g^{10}_x,\,g^{01}_x,\, g^\swap_x,\, g^{\tI\up-}_x,\, g^{\tI\dw-}_x,\, g^{\ta\up+}_x,\, g^{\ta\dw+}_x\right\}.
\end{equation}

The transition sum rules \eqref{eq:SIS-E2-trans} and uniformization sum rules \eqref{eq:SIS-E2-uni00} and \eqref{eq:SIS-E2-uni11} for the states $(0,0)$ and $(1,1)$ remain unchanged. Denoting the new vertex rates by $\nu^{10}_x$ and $\nu^{01}_x$, the adapted uniformization sum rules for the states $(1,0)$ and $(0,1)$ read
\begin{subequations}\label{eq:SIS-E3-uni10}
\begin{gather}
	\Gamma_x-\alpha_{i,j}=\nu^{10}_x+\nu^{\ta\dw+}_x+\nu^{\tI\up-}_x,\\
	\Gamma_x-\alpha_{j,i}=\nu^{01}_x+\nu^{\ta\up+}_x+\nu^{\tI\dw-}_x.
\end{gather}
\end{subequations}

This gives a system of five linear equations for nine parameters. Choosing the uniformization rate $\Gamma_x$, $\psi:=\nu^\swap_x$, $\phi^\up:=\nu^{\tI\up-}_x$, $\phi^\dw:=\nu^{\tI\dw-}_x$ as the free parameters, the solution space is
\begin{equation}\label{eq:SIS-E3-sol}
	\Pmatrix{\Gamma_x\\ \nu^\taa_x \\ \nu^{10}_x \\ \nu^{01}_x \\ \nu^\swap_x \\ \nu^{\tI\up-}_x \\ \nu^{\tI\dw-}_x \\ \nu^{\ta\up+}_x \\ \nu^{\ta\dw+}_x}
	= \Pmatrix{\Gamma_x\\ \Gamma_x \\ \Gamma_x - \alpha_{i,j} - \alpha_{j,i} + \psi - \Delta\phi \\ \Gamma_x - \alpha_{i,j} - \alpha_{j,i} + \psi + \Delta\phi \\ \psi \\ \phi^\up \\ \phi^\dw \\ \alpha_{i,j} - \psi - \phi^\up \\ \alpha_{j,i} - \psi - \phi^\dw}
\end{equation}
with $\Delta\phi:=\nu^{\tI\up-}_x-\nu^{\tI\dw-}_x$ and
\begin{subequations}\label{eq:SIS-E3-solRange}
\begin{gather}
	\Gamma_x \geq \alpha_{i,j} + \alpha_{j,i} - \psi + |\Delta\phi|\quad\text{and}\\
	0 \leq \psi \leq \min\big(\alpha_{i,j} - \phi^\up, \alpha_{j,i} - \phi^\dw\big)
\end{gather}
\end{subequations}
following from the constraint that all rates need to be nonnegative.

Choosing the minimal uniformization rate $\Gamma_x=\alpha_{i,j} + \alpha_{j,i} - \psi + |\Delta\phi|$, one of the blocker vertices is removed ($\nu^{10}_x=0$ or $\nu^{01}_x=0$). Choosing the maximal $\psi=\min(\alpha_{i,j} - \phi^\up, \alpha_{j,i} - \phi^\dw)$ to increase trunk mobility, one of the source-free vertices is removed ($\nu^{\ta\up+}_x=0$ or $\nu^{\ta\dw+}_x=0$). Lastly, a pragmatic choice would be
\begin{equation}
	\phi^\up=q\alpha_{i,j}\quad\text{and}\quad
	\phi^\dw=q\alpha_{j,i}
\end{equation}
with a suitable $0<q<1$ (to be optimized depending on the network).

Assuming $\alpha_{i,j}\geq\alpha_{j,i}$ without loss of generality, these choices lead to
\begin{align}\nonumber
	&\left(\Gamma_x,\, \nu^\taa_x,\, \nu^{10}_x,\, \nu^{01}_x,\, \nu^\swap_x,\, \nu^{\tI\up-}_x,\, \nu^{\tI\dw-}_x,\, \nu^{\ta\up+}_x,\, \nu^{\ta\dw+}_x\right)\\
	&\textstyle \nonumber
	=\Big((1+q)\alpha_{i,j},\, (1+q)\alpha_{i,j},\, 0, 2q\Delta\alpha,\, (1-q)\alpha_{j,i},\\
	& \qquad q\alpha_{i,j},\, q\alpha_{j,i},\, (1-q)\Delta\alpha,\, 0 \Big),
	\label{eq:SIS-E3-sol-q}
\end{align}
where $\Delta\alpha:=\alpha_{i,j}-\alpha_{j,i}$.

\section{Trajectory constraints \& initialization for SIS models}\label{sec:BC}
Concerning macroscopic constraints and the construction of a valid initial trajectory $\vs[1]$ for the path Markov chain \eqref{eq:pathChain}, let us discuss the specific case of an SIS model, where we want only one infected node at the starting time $t=0$ (patient zero) and at least $N^\tI_{\min}$ infected nodes at the final time $t=T$. To sample paths $\vs$ under these constraints, the dynamic programming method described in Sec.~\ref{sec:BC-DP} and Appendix~\ref{appx:BC-DP} is generally applicable and efficient, and it can be used for many other choices of trajectory constraints. As dynamic programming requires more coding effort, we also describe two simpler approaches based on conditional locks and cluster-level rejection sampling in Secs.~\ref{sec:BC-locks} and \ref{sec:BC-clusterRejection}. Finally, Sec.~\ref{sec:BC-init} describes how to initialize the path Markov chain \eqref{eq:pathChain} with a valid trajectory $\vs[1]$ that obeys the boundary conditions.

Below, we use the following setup: After the cluster identification (Sec.~\ref{sec:clusterIdentify}), we determine how many nodes are already permanently locked to the infected state $1$ at times $t=0$ and $T$. For a valid trajectory $\vs$, these are
\begin{equation}
	M^\tI_\lock\in\{0,1\}\quad\text{and}\quad
	N^\tI_\lock\in\{0,N\},
\end{equation}
respectively. Let $\C_1,\dotsc,\C_K$ be the $K$ free clusters that intersect the $t=0$ or $t=T$ time slices, forming the (generally overlapping) collections of clusters 
\begin{equation}
	\K_0\ \ \text{and}\ \ \K_T\subseteq\{\C_1,\dotsc,\C_K\}.
\end{equation}
Lastly, let
\begin{equation}
	m^\tI_k(a_k)\quad\text{and}\quad
	n^\tI_k(a_k)
\end{equation}
be the numbers of infected nodes contributed by cluster $\C_k$ in state $a_k\in\{0,1\}$ at times $t=0$ and $T$, respectively, with the mean $\bar{n}^\tI_k:=[n^\tI_k(0)+n^\tI_k(1)]/2$.

\subsection{Conditional locks}\label{sec:BC-locks}
A simple approach to enforce the boundary conditions are conditional locks on free clusters $\C_k\in\K_0$ that intersect $t=0$ and those $\C_k\in\K_T$ that intersect $t=T$.

For $t=0$, there are two cases: Either only one node is infected (patient zero) in $\vs$, or multiple nodes are infected. In the first case, for the cluster update, we freeze the $t=0$ states of all susceptible nodes except for the nearest neighbors of patient zero. In the second case, we freeze all susceptible nodes. When averaging over trajectories $\vs$ to estimate observables etc., we only take into account the trajectories $\vs$ with a single infected node at $t=0$.

For $t=T$, if $\sum_i \sigma_i^T\geq N^\tI_{\min}$, we don't freeze any node. If $\sum_i \sigma_i^T<N^\tI_{\min}$, we freeze the $t=T$ states of all nodes that are infected in the current trajectory $\vs$ before the update. When estimating observables, we only include the paths $\vs$ with at least $N^\tI_{\min}$ infected nodes at $t=T$.

This approach has two drawbacks: (i) A minor concern is that it breaks strict detailed balance \eqref{eq:detail_balance}. As a kind of boundary effect, this usually does not lead to observable errors. (ii) It reduces the efficiency of the cluster updates as we produce trajectories that are ultimately disregarded, and the locks generally increase trajectory autocorrelations.

\subsection{Cluster-level rejection sampling}\label{sec:BC-clusterRejection}
An alternative approach that resolves the problems of conditional locks is cluster-level rejection sampling. The idea is to satisfy the $t=0$ constraint by construction and use coin flips for the remaining free $t=T$ clusters. Because every valid combination $(a_1,\dotsc,a_K)$ of free clusters has equal statistical weight (before applying the constraints), we just need to ensure our proposal method samples uniformly from the subspace of trajectories that satisfy the patient-zero constraint.

Because background vertices leave the $t=0$ states of many nodes entirely decoupled from the epidemic, $\K_0$ is generally large, making naive random assignment computationally impossible as the chance of randomly drawing exactly one infected node vanishes exponentially. Instead, we can satisfy the $t=0$ constraint by exact construction. If $M^\tI_\lock = 1$, we fix all clusters in $\K_0$ to the state that yields 0 infected nodes at $t=0$. If $M^\tI_\lock = 0$, we uniformly randomly select exactly one cluster from $\K_0$ among those capable of providing exactly one infected node at $t=0$ (patient zero), fix its state accordingly, and set all remaining clusters in $\K_0$ to the state that yields 0 infected nodes at $t=0$. This exactly satisfies the initial boundary condition and inherently locks the state of any spanning clusters that intersect both boundaries \footnote{Note that the probability of free clusters that span the entire time interval from $[0,T]$ decreases exponentially in the final time $T$, because of the finite insertion rate of background graph vertices.}.

For the remaining free clusters $\C_k\in \K_T \setminus \K_0$, we randomly assign a state $a_k \in \{0,1\}$ with equal probability 1/2 each.
If the resulting total number of infected nodes at $t=T$ does \emph{not} meet the threshold $N^\tI_{\min}$, we discard the entire proposed assignment $(a_1,\dotsc,a_K)$, including the choice of patient zero, and repeat the process. This guarantees that we sample the valid joint configurations evenly, strictly obeying detailed balance \eqref{eq:detail_balance}.

Is this approach efficient? To satisfy the boundary condition $\sum_i \sigma_i^{\prime T} \geq N^\tI_{\min}$, the free clusters must collectively provide at least
\begin{equation}
	\Delta N^\tI = \max\left(0, N^\tI_{\min} - N^\tI_\lock\right)
\end{equation}
infected nodes. If the ratio $\Delta r:=\Delta N^\tI/ \sum_k \bar{n}^\tI_k$ is small, the random assignment of the states $a_k$ will succeed with high probability. This follows from the concentration of measure phenomenon and, in particular, Hoeffding's inequality \cite{Hoeffding1963-58,Boucheron2013}. The smallness of $\Delta r$ depends on several parameters, e.g., $N^\tI_{\min}/N$, the recovery rates $\gamma_i$, the insertion rates of the single-node graph vertices, the insertion rates of edge vertices that can add infections, and the node degrees.

Note that this approach would usually fail when using the edge vertex set \textsc{E1} in simulations of SIS models. Upstream lock avalanches would tend to freeze patient zero as discussed in Sec.~\ref{sec:SIS-E1-analysis}. The graph vertices $g^\swap_x$, $g^{\ta\up+}_x$, and $g^{\ta\dw+}_x$ of edge vertex set \textsc{E2} resolve this issue; see Sec.~\ref{sec:SIS-E2}.

\subsection{Dynamic programming}\label{sec:BC-DP}
If $\Delta N^\tI$ is substantially larger than $\sum_k \bar{n}^\tI_k$, the fraction of valid configurations out of the $2^K$ total possibilities becomes vanishingly small such that surpassing $\Delta N^\tI$ has low probability, and cluster-level rejection will stall.

Instead, we can exactly sample the valid combinations in $\O(KN_\text{free})$ time, where $N_\text{free}$ is the total number of nodes contained in the free clusters at time $T$: Build a $K\times 2\times N_\text{free}$ dynamic programming table $Z$ (similar to the Knapsack/subset-sum problems \cite{Martello1990,Kellerer2004}), where $Z_k(m,n)$ with $m\in\{0,1\}$ counts the number of valid state combinations $(a_1,\dotsc,a_k)$ for clusters $\C_1,\dotsc,\C_k$ that result in exactly $m=\sum_{k'=1}^k m^\tI_{k'}(a_{k'})$ infected nodes at $t=0$ and $n=\sum_{k'=1}^k n^\tI_{k'}(a_{k'})$ infected nodes at $t=T$ in those clusters. Once the table is built, do a single backward pass, probabilistically assign the cluster-$\C_K$ state $a_K\in\{0,1\}$ based on the exact combinatorial weight of the remaining paths that reach the threshold $\Delta N^\tI$ and one patient zero. Then move to $a_{K-1}$, and so on. Details are provided in Appendix~\ref{appx:BC-DP}.

\subsection{Initialization}\label{sec:BC-init}
To initialize a CPMC simulation with a single patient-zero and outbreak-size threshold $N^\tI_{\min}$, assuming a connected network, we can start from a trajectory $\vs[1]$, where a random node is infected for the entire time interval $t\in[0,T]$ and all other nodes are susceptible. Starting with $n=1$, we iterate the following:
Given the current trajectory $\vs[n]$, do a cluster update with outbreak threshold $\tilde{N}^\tI_{\min}=1+\sum_i \sigma_i^T[n]$ to generate a trajectory $\vs[n+1]$ with at least one additional infected node at time $T$. Continue this process until reaching a trajectory $\vs[n+1]$ that surpasses the desired infection threshold $N^\tI_{\min}$. This trajectory $\vs[n+1]$ can then be used as the first sample in the actual CPMC Markov chain \eqref{eq:pathChain}.

\section{Computation costs}\label{sec:costs}
The computational efficiency of CPMC depends on the time complexity of a single cluster update $\vs[n] \to \vg \to \vs[n+1]$ (Sec.~\ref{sec:clusterUpdate}) and autocorrelations in the resulting path Markov chain \eqref{eq:pathChain}.

Generating the intermediate graph $\vg$ involves processing the $\Np_\vs$ physical events and inserting $\Nb_{\vg,\vs}$ background vertices on intervals where the state is constant through a Poisson point process as discussed in Sec.~\ref{sec:grapConstruct}. The computational cost generally is
\begin{equation}\label{eq:cost-graph}\textstyle
	\O(\Np_\vs+\Nb_{\vg,\vs})=\O\big(\int_0^T\ud t\,\Gamma^t\big),
\end{equation}
where $\Gamma^t$ is the total uniformization rate $\Gamma^t$ [Eq.~\eqref{eq:def-Lambda-total}].

Translating the generated graph $\vg$ into spacetime clusters $\C_k$ requires applying the union-find (disjoint-set) algorithm across the trajectory as discussed in Sec.~\ref{sec:clusterIdentify}. When using path compression (flattening the cluster tree every time we search for a root) and union-by-rank (always attaching the smaller cluster tree to the root of the larger tree), the computational complexity scales almost linearly with the total number of graph vertices \eqref{eq:cost-graph}.

Enforcing macroscopic constraints without path rejection introduces a small overhead. For instance, enforcing the boundary condition of $N^\tI_{\min}$ infected nodes at time $t=T$ using dynamic programming requires building the combinatorial table $Z_k(m,n)$ described in Sec.~\ref{sec:BC} and Appendix~\ref{appx:BC-DP}. This process operates in $\O(K N_\text{free})$ time, where $K$ is the number of free clusters intersecting the final time $t=T$ and $N_\text{free}$ is the total number of nodes contained within those clusters at time $T$.

\begin{example}
Consider simulating an SIS model with recovery rates $\gamma_i$ and infection rates $\alpha_{i,j}$ using the single-node vertex set \textup{\textsc{S1}} and edge-vertex set \textup{\textsc{E2}} from Sec.~\ref{sec:SIS-vertexSets}. According to Eqs.~\eqref{eq:def-Lambda-total} and \eqref{eq:cost-graph} the time complexity per cluster update scales as
\begin{equation}\textstyle
	\O\big(\int_0^T\ud t\,\Gamma^t\big)=\O\big(T\sum_i\gamma_i\big)+\O\big(T\sum_{i,j}\alpha_{i,j}\big).
\end{equation}
\end{example}

Autocorrelation times are much more difficult to predict and generally depend on network structure, the distribution of event rates, as well as the specific trajectory constraint $C(\vs)$. They can also crucially depend on the chosen sets of graph vertices and associated vertex rates. As discussed  for SIS models in Secs.~\ref{sec:SIS-E1-analysis} and \ref{sec:SIS-E2}, we certainly want to avoid any (upstream and downstream) lock avalanches that lead to small numbers of free clusters. Lowering the computation cost per cluster update can increase autocorrelation times in the path Markov chain \eqref{eq:pathChain} and increase the net computation time to generate trajectories that are (approximately) statistically independent. 

\section{Parallelization strategies}\label{sec:parallel}
To scale CPMC for massive networks, the algorithm can be parallelized in multiple ways:

\Emph{Independent Markov chains (trivial parallelization).}~--
Because CPMC generates a Markov chain of full-system trajectories, multiple statistically independent runs can be trivially distributed across separate CPU cores. The results of the independent Markov chains can be concatenated to reduce the statistical variance of the sampled observables. A limitation in this respect is that every Markov chain requires a warm-up phase (a.k.a.\ burn-in period). With the latter typically comprising 20\% to 40\% of the generated samples \cite{Gelman2013}, the possible gains are limited.

\Emph{Temporal decomposition for graph generation.}~--
The graph generation can be easily parallelized. For the assignment of graph vertices to the state-changing events in the current trajectory $\vs$, we can simply assign equal-sized groups of vertices to every thread. For the insertion of background graph vertices, we can divide the time axis  $[0,T]$ into non-overlapping equal-sized intervals. Different threads can independently generate background graph vertices $g_y$ for their assigned time windows.

\Emph{Concurrent union-find data structures.}~--
While a standard union-find algorithm is inherently sequential, concurrent lock-free disjoint-set data structures can be implemented \cite{Anderson1991-23,Alistarh2019-153,Jayanti2021-34}. By distributing the nodes $i\in\{1,\dotsc,N\}$ across parallel threads, the initial sequence of constant-state segments and independent tree roots can be generated simultaneously. Threads can then safely execute atomic ``compare-and-swap'' unifications for the edge vertices $g_{(t,i,j)}$ to accelerate the cluster identification.

\Emph{Parallel tempering (replica exchange).}~--
For large networks with rigid topological constraints, we can employ parallel tempering, simulating multiple, coupled Markov chains running simultaneously on separate processors \cite{Note4}. Autocorrelation times can be reduced by swapping configurations between chains running at different structural parameter regimes, e.g., varying the strictness of the trajectory constraint $C(\vs)$, model parameters like $\gamma_i$ and $\alpha_{i,j}$ in SIS dynamics, and tuning algorithmic parameters like $\psi$ and $\phi$ for the vertex rates in edge vertex set \textsc{E2} of Sec.~\ref{sec:SIS-E2}.

\section{Validation against exact solutions}\label{sec:validate}
To demonstrate and validate the method for large outbreaks on big SIS networks, simulation results of CPMC and the Gillespie method are compared in Ref.~\cite{Sun2026_08} for regimes, where the rejection rate of the Gillespie method (due to the outbreak-size constraint) is high but still manageable.

Here we validate CPMC by comparing to the exact solution of the master equation for two small four-node SIS networks.

\Emph{Four-node ring.}~--
For a homogeneous ring network of $N=4$ nodes, the  cyclic translation invariance allows us to reduce the $2^4 = 16$ distinct microstates down to the $6$ macrostates
\begin{subequations}\label{eq:ringStates}
\begin{gather}
	s_0=(0,0,0,0),\quad s_1=(1,0,0,0),\\
	s_2=(1,1,0,0),\quad s_{\tilde{2}}=(1,0,1,0),\\
	s_3=(1,1,1,0),\quad s_4=(1,1,1,1),
\end{gather}
\end{subequations}
which are representatives for the 1, 4, 4, 2, 4, 1 microstates that they are related to by translations.

Let $p_i(t)$ denote the probability of the system being in macrostate $s_i$ at time $t$. By analyzing the number of active edges and possible recovery events for each macrostate, we obtain the following transition rates:
\begin{itemize}
 \item From $s_1$: Recovers to $s_0$ (rate $\gamma$). Infects an adjacent neighbor to reach $s_2$ (rate $2\alpha$).
 \item From $s_2$: Recovers to $s_1$ (rate $2\gamma$). Infects a susceptible node to reach $s_3$ (rate $2\alpha$).
 \item From $s_{\tilde{2}}$: Recovers to $s_1$ (rate $2\gamma$). Both susceptible nodes are flanked by two infected nodes, yielding $4$ active edges to reach $s_3$ (rate $4\alpha$).
 \item From $s_3$: Recovers an ``end'' node to reach $s_2$ (rate $2\gamma$). Recovers the ``middle'' node to reach $s_{\tilde{2}}$ (rate $\gamma$). Infects the final node to reach $s_4$ (rate $2\alpha$).
 \item From $s_4$: Recovers to $s_3$ (rate $4\gamma$).
\end{itemize}
\begin{figure}[t]
\label{fig:ring}
\includegraphics[width=0.97\columnwidth]{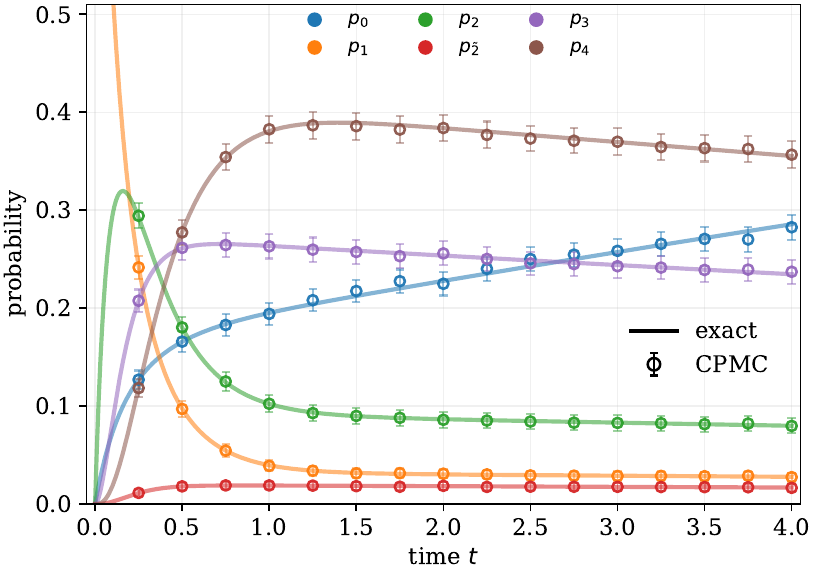}
\caption{\Emph{Homogeneous four-node SIS ring network with $\gamma=1$ and $\alpha=3$:} Probabilities $p_i(t)$ for finding the system in macrostates $s_i$ [Eq.~\eqref{eq:ringStates}] at time $t$, when starting in one of the four $s_1$ states.}
\end{figure}

This yields the following system of coupled differential equations:
\begin{subequations}
\begin{align}
	\partial_t p_0 &= \gamma p_1 \\
	\partial_t p_1 &= -(2\alpha + \gamma)p_1 + 2\gamma p_2 + 2\gamma p_{\tilde{2}} \\
	\partial_t p_2 &= 2\alpha p_1 - (2\alpha + 2\gamma)p_2 + 2\gamma p_3 \\
	\partial_t p_{\tilde{2}} &= -(4\alpha + 2\gamma)p_{\tilde{2}} + \gamma p_3 \\
	\partial_t p_3 &= 2\alpha p_{2} + 4\alpha p_{\tilde{2}} - (2\alpha + 3\gamma)p_3 + 4\gamma p_4 \\
	\partial_t p_4 &= 2\alpha p_3 - 4\gamma p_4
\end{align}
\end{subequations}
which can be written and solved in matrix form as
\begin{equation}\label{eq:exactSol}
	\partial_t \vec{p}(t) = A\, \vec{p}(t)\quad\text{such that}\quad
	\vec{p}(t) = e^{At }\vec{p}(0).
\end{equation}
Choosing $\gamma=1$, $\alpha=3$, and $s_1$ as the initial state, the data in Fig.~\ref{fig:ring} show excellent agreement between the CPMC simulation and the exact solution.

\Emph{Four-node diamond.}~--
This network comprises two hub nodes $i=1,2$ that are connected to all other nodes, and two leaf nodes $i=3,4$ which are connected to the hubs but not directly to each other. We choose all infection rates on edges and recovery rates on nodes to be equal ($\alpha$ and $\gamma$). All configurations with the same numbers $n_h,n_\ell\in\{0,1,2\}$ of infected hub and leaf nodes are dynamically equivalent. This restricted permutation symmetry allows us to reduce the $2^4 = 16$ distinct microstates down to $3\cdot 3=9$ macrostates which evolve as follows:
\begin{itemize}
 \item Hub recovery ($n_h \to n_h-1$): The rate is $n_h\gamma$.
 \item Leaf recovery ($n_\ell \to n_\ell-1$): Rate is $n_\ell\gamma$.
 \item Hub infection ($n_h \to n_h+1$): A susceptible hub is connected to the other hub and both leaves. If the other hub is infected, it contributes $\alpha$. Each infected leaf contributes $\alpha$. The total rate for the $2-n_h$ susceptible hubs is $(2-n_h)(n_h + n_\ell)\alpha$.
 \item Leaf infection ($n_\ell \to n_\ell+1$): A susceptible leaf is connected only to the hubs. Each infected hub contributes $\alpha$. The total rate for the $2-n_\ell$ susceptible leaves is $(2-n_\ell)n_h\alpha$.
\end{itemize}
\begin{figure}[t]
\label{fig:diamond2-2}
\includegraphics[width=0.97\columnwidth]{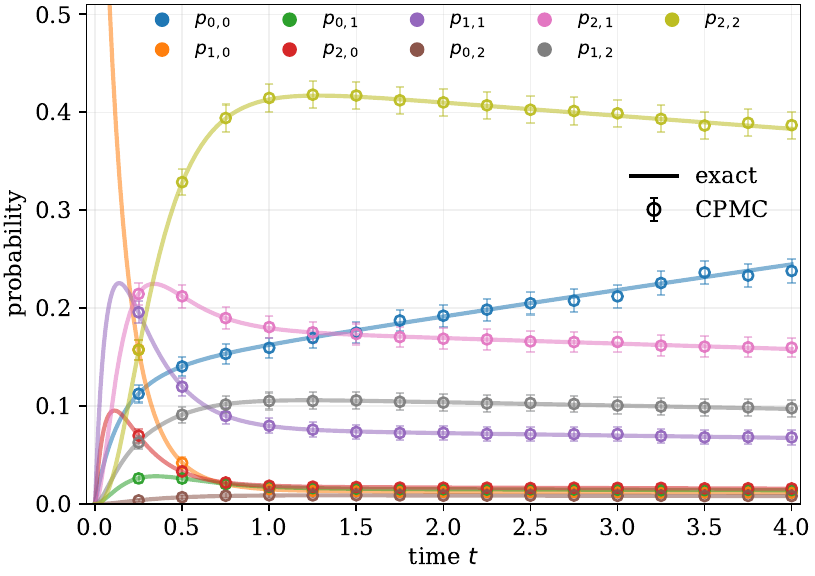}
\caption{\Emph{Four-node diamond SIS network with two hub nodes and two leaf nodes, $\gamma=1$ and $\alpha=5/2$:} Probabilities $p_{n_h,n_\ell}(t)$ for finding the system with $(n_h,n_\ell)$ infected nodes hub and leaf nodes at time $t$, when starting in an $(n_h,n_\ell)=(1,0)$ state.}
\end{figure}

Using these macroscopic rates, the master equations governing the probability $p_{n_h,n_\ell}=p_{n_h,n_\ell}(t)$ of finding the system with $(n_h,n_\ell)$ infected nodes at time $t$ are
\begin{align*}
	\partial_t p_{0,0} &= \gamma p_{1,0} + \gamma p_{0,1}, \\
	\partial_t p_{1,0} &= -(3\alpha + \gamma)p_{1,0} + 2\gamma p_{2,0} + \gamma p_{1,1}, \\
	\partial_t p_{0,1} &= -(2\alpha + \gamma)p_{0,1} + \gamma p_{1,1} + 2\gamma p_{0,2}, \\
	\partial_t p_{2,0} &= \alpha p_{1,0} - (4\alpha + 2\gamma)p_{2,0} + \gamma p_{2,1}, \\
	\partial_t p_{1,1} &= 2\alpha p_{1,0} + 2\alpha p_{0,1} - (3\alpha + 2\gamma)p_{1,1}\\
	&\quad+ 2\gamma p_{2,1} + 2\gamma p_{1,2}, \\
	\partial_t p_{0,2} &= -(4\alpha + 2\gamma)p_{0,2} + \gamma p_{1,2}, \\
	\partial_t p_{2,1} &= 4\alpha p_{2,0} + 2\alpha p_{1,1} - (2\alpha + 3\gamma)p_{2,1} + 2\gamma p_{2,2}, \\
	\partial_t p_{1,2} &= \alpha p_{1,1} + 4\alpha p_{0,2} - (3\alpha + 3\gamma)p_{1,2} + 2\gamma p_{2,2}, \\
	\partial_t p_{2,2} &= 2\alpha p_{2,1} + 3\alpha p_{1,2} - 4\gamma p_{2,2},
\end{align*}
which can again be written and solved in matrix form according to Eq.~\eqref{eq:exactSol}.
Choosing $\gamma=1$, $\alpha=5/2$, and $(n_h,n_\ell)=(1,0)$ as the initial state, the data in Fig.~\ref{fig:diamond2-2} show excellent agreement between the CPMC simulation and the exact solution.

Note that, in these validation numerics, we have not imposed any constraint on the final state to simplify the exact solution.

\section{Discussion}\label{sec:discuss}
In this manuscript, we have established the rigorous mathematical foundations and the algorithmic framework for conditional-path Monte Carlo (CPMC), complementing the broader introduction of the method in Ref.~\cite{Sun2026_08}.

The individual-based simulation of rare macroscopic events in stochastic dynamics on large networks represents a considerable challenge. As outlined in Sec.~\ref{sec:intro}, techniques like the traditional SSA \cite{Gillespie1977-81}, wSSA \cite{Kuwahara2008-129,Gillespie2009-130}, splitting methods 
\cite{Allen2009-21,VillenAltamirano2002-13,Zuckerman2017-46,Aristoff2018-52}, and TPS \cite{Bolhuis2002-53,Dellago2002-1,Bolhuis2021-4} routinely succumb to catastrophic rejection rates, kinetic trapping, path degeneracy, critical slowing down, or genealogical correlations when faced with large heterogeneous networks, rigid topological bottlenecks, or complex boundary conditions. 
The CPMC method aims to circumvent these structural vulnerabilities by abandoning forward-time integration and local trajectory updates. Instead, CPMC employs cluster updates that map a given full-system trajectory to an intermediate graph configuration, decomposing spacetime into non-local clusters, which are then updated through (largely independent) cluster-state flipping.

\Emph{Methodological advances.}~-- 
The core theoretical contribution is the formulation of the joint path-graph probability weights and the derivation of the local transition and uniformization sum rules that guarantee strict detailed balance for the trajectory Markov chain. During the construction of corresponding graph vertex sets for SIS models, we found that a careless choice can lead to upstream (or downstream) lock avalanches in the generated graphs, reducing the size or number of flippable clusters and demobilizing the epidemic trunk. This can lead to considerable autocorrelations. By systematically constructing and optimizing the edge vertex sets (from \textsc{E1} to \textsc{E2} and \textsc{E3}), we demonstrated how specific graph vertices such as the swap and asymmetry-nullifying vertices ($g^\swap_x$, $g^{10}_x$, and $g^{01}_x$) can completely unlock trajectory mobility, ensuring ergodicity and accommodating highly asymmetric infection rates without sacrificing computational efficiency.

Importantly, CPMC resolves the severe inefficiencies of SSA-based techniques associated with macroscopic trajectory constraints. Enforcing conditions such as a single patient zero at $t=0$ alongside a large outbreak threshold at $t=T$ in SIS models typically destroys the acceptance rates of stochastic algorithms. By constructing a truncated dynamic programming table \cite{Martello1990,Kellerer2004} for the free clusters that can affect the macroscopic trajectory constraints, CPMC can efficiently sample the compatible trajectories without rejection and with exact detailed balance.

\Emph{Generalizability and applications.}~-- 
While this paper exclusively employed the SIS model for instructive clarity in examples, the CPMC method and the logic behind the vertex set construction extend naturally to general Markovian stochastic dynamics. The ability of CPMC to bypass path degeneracy and mutate non-local spacetime clusters makes it exceptionally well-suited for a wide array of complex systems where rare combinations of local processes trigger massive systemic shifts.
Beyond epidemiology, this framework holds immediate potential for sampling the rare transition pathways of nucleation in kinetic spin models \cite{Binder1987-50,Rikvold1994-49} and overcoming the severe kinetic trapping inherent to the space-time thermodynamics of glassy systems \cite{Garrahan2007-98,Chandler2010-61}. CPMC can be used to analyze cascading failures in IT and financial networks \cite{Albert2000-406,Acemoglu2015-105}, targeted lateral malware propagation \cite{Newman2002-66}, and rapid shifts in societal opinions \cite{Castellano2009-81,Watts2002-99}. In these highly heterogeneous systems, macroscopic metrics are often insufficient for an in-depth investigation, understanding, and control of the complex dynamics. CPMC enables an efficient individual-based simulation and risk factor analysis for important rare events.

\Emph{Future directions.}~-- 
Several avenues for future methodological developments remain. First, adapting the graph vertex sets to other canonical epidemiological models, such as the susceptible-infected-recovered (SIR) or susceptible-exposed-infected-recovered (SEIR) models \cite{PastorSatorras2015-87}, will involve absorbing states and non-reversible node transitions. Second, it will be interesting to apply CPMC to systems with time-dependent event rates and to temporal networks, where the edges themselves appear and disappear dynamically \cite{Holme2012-519}. While the presented general formulation and SIS graph vertex sets are applicable for these cases, the insertion process for background vertices will become more complex, and details for an efficient implementation need to be worked out. Finally, while we have derived the solution space for vertex rates in suitable SIS graph vertex sets (e.g., the parameters $\psi, \phi$, and $q$ in the \textsc{E2} and \textsc{E3} edge sets), the identification of (model-dependent) parameter choices that minimize autocorrelation times remains an open optimization problem. Combining CPMC with parallel tempering \cite{Note4} and adaptive machine learning algorithms to actively tune vertex rates during the burn-in phase could further push the boundaries of rare-event simulation for massive, real-world networks.

\begin{acknowledgments}
We gratefully acknowledge discussions with Caterina de Bacco, Yuna Jang, Jianfeng Lu, James Moody, Charles Nunn, Joshua Socolar, and participants of the international workshop ``Quantitative Methods for Dynamics on Networks'' 2024 in Los Alamos (organized by the CNLS at Los Alamos National Laboratory), as well as support by the U.S.\ National Science Foundation through grant no.\ DMS-2344576 and by the Duke Population Research Center (DPRC) through the U.S.\ NICHD grant no.\ P2C-HD0065563.
\end{acknowledgments}

\onecolumngrid
\appendix
\section{Proof of the joint-weight marginalization condition}\label{appx:Jmargin-g}
In this appendix, we shall prove the marginalization condition $P(\vs)= \sum'_{\vg\ni\vs} J(\vs,\vg)$ [Eq.~\eqref{eq:Jmargin-g}], stating that the path probability density $P(\vs)$ [Eq.~\eqref{eq:def-P}] is recovered when summing the joint path-graph weights $J(\vs,\vg)$ [Eq.~\eqref{eq:def-J}] over all graphs $\vg$ compatible with $\vs$, where the ``sum'' $\sum'_{\vg\ni\vs}$ is actually the path integral \eqref{eq:def-graphSum}.

Let us split the product $\prod_{z\in\vg}\nu_z(g_z)\Delta(\vs_z,g_z)$ from Eq.~\eqref{eq:def-J} into the product $\prod_{x\in\vs}\nu_x(g_x)\Delta(\vs_x,g_x)$ over the locations $\{x\}$ of all $\Np=\Np_\vs$ physical vertices and the product $\prod_{y}\nu_y(g_y)\Delta(\vs_y,g_y)$ over the locations $\{y\}$ of all $\Nb=\Nb_{\vg,\vs}$ background vertices with respect to trajectory $\vs$. Applying the transition sum rule \eqref{eq:nu-transitionRule} for the graph vertex rates $\nu$ in the physical-vertex product, we have
\begin{equation}\label{eq:J-sump-g_simplify-x}
	\sum_{g_{x_1},\dotsc,g_{x_\Np}} \prod_{x\in\vs}\nu_x(g_x)\Delta(\vs_x,g_x)
	\stackrel{\eqref{eq:nu-transitionRule}}{=} \prod_{x\in\vs}\omega_x(\vs_x).
\end{equation}
Under the path integral, the background-vertex product evaluates to
\begin{align}\nonumber
	&\sum_{\Nb=0}^\infty \frac{1}{\Nb!}\int_0^T \ud t_1\dots\ud t_\Nb 
	\sum_{\vy_1,\dotsc,\vy_\Nb}\,\sum_{g_{y_1}\in\G_{\vy_1}^{t_1}}\dots\sum_{g_{y_\Nb}\in\G_{\vy_\Nb}^{t_\Nb}}
	\prod_{m=1}^\Nb \nu_{y_m}(g_{y_m})\Delta(\vs_{y_m},g_{y_m}) \\\nonumber
	&=\sum_{\Nb=0}^\infty \frac{1}{\Nb!}\Big(\int_0^T \ud t\sum_\vy \sum_{g_y\in\G_y}\nu_y(g_y)\Delta(\vs_y,g_y)\Big)^\Nb
	 =\exp\Big(\int_0^T \ud t\sum_\vy \sum_{g_y\in\G_y}\nu_y(g_y)\Delta(\vs_y,g_y)\Big)\\
	&=\exp\Big(\int_0^T \ud t\sum_\vy \sum_{g_y\in\G_y}\nu_y(g_y)\Delta(\vs_y^{\const},g_y)\Big)
	 \stackrel{\eqref{eq:nu-uniformSumRule}}{=}
	  \exp\Big(\int_0^T \ud t\sum_\vy \big[\Gamma_y - \Lambda_y(\vs_\vy^t)\big]\Big)
	 =e^{\int_0^T\ud t\,[\Gamma^t-\Lambda^t(\vs^t)]}.
	\label{eq:J-sump-g_simplify-y}
\end{align}
For the second line, recall that $\G_y\equiv\G_\vy^t$. In the third line, we have first used that state-changing events have measure zero on the time interval $[0,T]$ such that they do not contribute in the integral over $t$ and we can replace all local state dynamics $\vs_y$ by the null-event $\vs_y^{\const}$ [Eq.~\eqref{eq:def-nonEvent}]. We have then employed the uniformization sum rule \eqref{eq:nu-uniformSumRule}.

In conjunction, Eqs.~\eqref{eq:def-P}, \eqref{eq:def-J}, \eqref{eq:def-graphSum}, \eqref{eq:J-sump-g_simplify-x}, and \eqref{eq:J-sump-g_simplify-y} prove the marginalization condition
\begin{equation}\label{eq:Jmarginal-step2}
	\sump_{\vg\ni\vs} J(\vs,\vg) =
	C(\vs) \Big(\prod_{x\in\vs}\omega_x(\vs_x)\Big) e^{-\int_0^T\ud t\,\Lambda^t(\vs^t)}
	=P(\vs).
\end{equation}

\section{Complete edge vertex set for SIS models}\label{appx:SIS-completeEdgeSet}
Section~\ref{sec:SIS-E2} contained the claim that
\begin{subequations}\label{eq:SIS-E-Ivertices}
\begin{alignat}{9}
	&g^{\ta\up-}_x	&&=(a,0\to a,a),\qquad &&g^{\tI\up-}_x	&&=(1,0\to 1,a),\qquad &&g^{\tI\up+}_x	&&=(1,a\to 1,1),\qquad &&g^{\ta\up+}_x	&&=(a,\bar{a}\to a,1),\\
	&g^{\ta\dw-}_x	&&=(0,a\to a,a),       &&g^{\tI\dw-}_x	&&=(0,1\to a,1),       &&g^{\tI\dw+}_x	&&=(a,1\to 1,1),       &&g^{\ta\dw+}_x	&&=(\bar{a},a\to 1,a),\\
	&g^\swap_x		&&=(a,\bar{a}\to 1,1)
\end{alignat}
\end{subequations}
are all edge vertices that are compatible with an infection event and feature at least one free variable $a\in\{0,1\}$ with $\bar{a}$ denoting the negation of $a$.

To prove this, consider a vertex $g_x$ that matches the infection event $(1,0\to 1,1)$ when evaluated at $a=1$. The vertex must evaluate to one of the remaining physically valid edge events when $a=0$:
\begin{enumerate}
 \item Mapping to $(0,0\to 0,0)$ implies $g_x=(a,0\to a,a)=g^{\ta\up-}_x$.
 \item Mapping to $(1,0\to 1,0)$ implies $g_x=(1,0\to 1,a)=g^{\tI\up-}_x$.
 \item Mapping to $(0,1\to 0,1)$ implies $g_x=(a,\bar{a}\to a,1)=g^{\ta\up+}_x$.
 \item Mapping to $(1,1\to 1,1)$ implies $g_x=(1,a\to 1,1)=g^{\tI\up+}_x$.
 \item Mapping to $(0,1\to 1,1)$ implies $g_x=(a,\bar{a}\to 1,1)=g^\swap_x$.
\end{enumerate}
Then considering the second option, a vertex $g_x$ that matches the infection event $(0,1\to 1,1)$ when evaluated at $a=1$, we find the remaining four vertices $g^{\ta\dw-}_x$, $g^{\tI\dw-}_x$, $g^{\ta\dw+}_x$, and $g^{\tI\dw+}_x$.

\section{Dynamic programming for the SIS-model boundary conditions}\label{appx:BC-DP}
Using the same notations as in Sec.~\ref{sec:BC}, this appendix describes in detail how we can use dynamic programming to exactly sample states $(a_1,\dotsc,a_K)$ of the $K$ free cluster that simultaneously satisfy the constraints at both boundaries: having exactly one patient zero at $t=0$ and at least $N^\tI_{\min}$ infected nodes at the final time $t=T$.

The following dynamic programming approach finds valid states of the free clusters in $\O(KN_\text{free})$ time and obeys strict detailed balance. It is split into a forward counting phase and a backward sampling phase and expands on the classic Knapsack/subset-sum algorithm \cite{Martello1990,Kellerer2004} by explicitly coupling the two boundary conditions into a single truncated state space.

\subsection{Forward pass -- Building the combinatorial table}
Considering only the first $k$ clusters, we want to count the number of valid ways $Z_k(m,n)$ the clusters can be assigned states $(a_1,\dotsc,a_k)$ to achieve exactly $m=\sum_{k'=1}^{k} m^\tI_{k'}$ infected nodes at $t=0$ and $n=\sum_{k'=1}^{k} n^\tI_{k'}$ infected nodes at $t=T$. Due to the patient-zero constraint, we only need to consider $m\in\{0,1\}$ such that $Z$ is a $K\times 2\times N_\text{free}$ combinatorial table.

To initialize the table, we set $Z_0(0,0)=1$ and $Z_0(m,n)=0$ for all $n>0$ and $m\in\{0,1\}$. We then loop through the cluster number $k$ from $1$ to $K$ and, for each possible pair $(m,n)$, calculate the number of valid combinations. The combination of sums $(m,n)$ can be reached either by setting the $k$-th cluster to $a_k=0$, inheriting the combinations for sums $\big(m-m^\tI_k(0), n-n^\tI_k(0)\big)$, or by setting it to $a_k=1$, inheriting the combinations for sums $\big(m-m^\tI_k(1), n-n^\tI_k(1)\big)$, i.e.,
\begin{equation}
	Z_k(m,n) = Z_{k-1}\big(m-m^\tI_k(0), n-n^\tI_k(0)\big) + Z_{k-1}\big(m-m^\tI_k(1), n-n^\tI_k(1)\big),
\end{equation}
where $Z_k(m,n)\equiv 0$ if $m<0$ or $n<0$.

By the end of this pass, $Z_K(m,n)$ holds the exact number of ways the entire set of $K$ free clusters can simultaneously produce exactly $m$ infected nodes at $t=0$ and $n$ infected nodes at $t=T$.

\subsection{Backward pass -- Exact joint sampling}
Let us define a helper function $S_k(m,n)$ which gives the total number of ways in which the first $k$ clusters $\C_1,\dotsc,\C_k$ can produce $m$ infected nodes at $t=0$ and \emph{at least} $n$ infected nodes at $t=0$:
\begin{equation}
	S_k(m,n) := \sum_{n'=n}^{N_{\text{free}}} Z_k(m,n').
\end{equation}

Now, we choose cluster states in the order $a_K, a_{K-1}, \dotsc, a_1$ one by one, sampling each state based on the exact marginal probability that it leads to a valid final trajectory. Starting with the target thresholds $n\leftarrow \Delta N^\tI$ and $m\leftarrow 1- M^\tI_\lock$, we iterate backward from $k = K$ down to $1$:
\begin{itemize}
 \item 
 Calculate the probabilities for the states $a_k\in\{0,1\}$ of cluster $\C_k$: For state $a_k$, cluster $\C_k$ contributes $m^\tI_k(a_k)$ nodes at $t=0$ and $n^\tI_k(a_k)$ nodes at $t=T$. Hence, the remaining $k-1$ clusters only need to provide
 \begin{subequations}
 \begin{alignat}{3}
	m'(a_k)&:=\max\big(0, m-m^\tI_k(a_k)\big)\quad &&\text{nodes at}\ \ t=0\quad\text{and}\\
	n'(a_k)&:=\max\big(0, n-n^\tI_k(a_k)\big)\quad &&\text{nodes at}\ \ t=T.
 \end{alignat}
 \end{subequations}
 The corresponding probability is exactly the number of valid combinations with state $a_k$, divided by the total number of valid combinations currently available:
 \begin{equation}
	p(a_k) = \frac{S_{k-1}\big(m'(a_k),n'(a_k)\big)}{S_k(m,n)}.
 \end{equation}
 \item
 Randomly select $a_k$ with probability $p(a_k)$, and update the requirements $m\leftarrow m'(a_k)$ and $n\leftarrow n'(a_k)$.
 \item
 Continue with $k\leftarrow k-1$.
\end{itemize}
In this way, every valid configuration $(a_1,\dotsc,a_K)$ satisfying both boundary conditions is chosen with equal probability and zero rejections.

As a minor optimization, we could take into account that cluster $\C_k$ necessarily contributes at least $\min\big(n^\tI_k(0),n^\tI_k(1)\big)$ infected nodes at $t=T$ and that we can increase the number of infected nodes at most by $\tilde{N}_{\text{free}}:=\sum_{k=1}^K\big|n^\tI_k(1)-n^\tI_k(0)\big|$ from the corresponding baseline. This reduces the $n$-dimension of the combinatorial table to $\tilde{N}_\text{free}$.

\end{document}